\newif\ifjournalstyle\IfFileExists{sn-jnl.cls}{\journalstyletrue\documentclass[a4paper,pdflatex,sn-mathphys-num]{sn-jnl}}{\documentclass[a4paper,10pt]{article}}
\usepackage{amsmath,amssymb,mathtools,booktabs,array,amsthm,float}\usepackage[protrusion=true,expansion=true]{microtype}
\ifjournalstyle\else\usepackage[a4paper,textwidth=144mm,textheight=225mm]{geometry}\usepackage[numbers,sort&compress]{natbib}\usepackage[colorlinks=true,linkcolor=blue,citecolor=blue,urlcolor=blue]{hyperref}\fi
\ifjournalstyle\theoremstyle{thmstyleone}\else\theoremstyle{plain}\fi
\newtheorem{theorem}{Theorem}[section]
\newtheorem{lemma}[theorem]{Lemma}
\newtheorem{proposition}[theorem]{Proposition}
\newtheorem{corollary}[theorem]{Corollary}
\ifjournalstyle\let\paperabstract\abstract\else\newcommand{\paperabstract}[1]{\begin{abstract}#1\end{abstract}}\fi
\ifjournalstyle\else\let\articletitle\title\renewcommand{\title}[2][]{\articletitle{#2}}\newcommand{\keywords}[1]{\par\noindent\textbf{Keywords}\enspace #1\par}\fi
\begin{document}
\hypersetup{pdftitle={Distribution of symmetric partial Boolean functions by query complexity and real degree}, pdfauthor={Guoliang Xu, Mingke Liu, Mengfan Wu, Haoran Zhang}}
\title[Distribution of symmetric partial Boolean functions]{Distribution of symmetric partial Boolean functions by query complexity and real degree}
\ifjournalstyle
  \author*[1]{\fnm{Guoliang} \sur{Xu}}\email{xu1guo2liang@foxmail.com}
  \author[2]{\fnm{Mingke} \sur{Liu}}
  \author[1]{\fnm{Mengfan} \sur{Wu}}
  \author[1]{\fnm{Haoran} \sur{Zhang}}
  \affil*[1]{\orgdiv{School of Software}, \orgname{Luoyang Normal University},
    \orgaddress{\city{Luoyang}, \postcode{471934}, \country{China}}}
  \affil[2]{\orgdiv{School of Artificial Intelligence}, \orgname{Luoyang Normal University},
    \orgaddress{\city{Luoyang}, \postcode{471934}, \country{China}}}
\else
  \author{Guoliang Xu\textsuperscript{1,*}, Mingke Liu\textsuperscript{2},
    Mengfan Wu\textsuperscript{1}, and Haoran Zhang\textsuperscript{1}\\[3pt]
    \small\textsuperscript{1} School of Software, Luoyang Normal University, Luoyang 471934, China\\
    \small\textsuperscript{2} School of Artificial Intelligence, Luoyang Normal University, Luoyang 471934, China\\
    \small\textsuperscript{*} Corresponding author:
      \href{mailto:xu1guo2liang@foxmail.com}{\nolinkurl{xu1guo2liang@foxmail.com}}}
  \date{}
  \maketitle
\fi
\paperabstract{Query complexity measures the minimum number of oracle calls needed to compute a function in a given model. A symmetric partial Boolean function on $n$ bits assigns a binary output to each permitted Hamming weight (number of ones). We count these functions by exact quantum query complexity, bounded-error randomized query complexity, and free real degree. The free real degree is the least degree of a real polynomial matching the specified labels. For each fixed integer $k\ge1$, exact quantum cost $k$ occurs for $\Theta_k(n^k)$ functions. With worst-case error bound $0<\epsilon<1/2$ also fixed, randomized cost $k$ occurs for $\Theta_{k,\epsilon}(2^{(1-\delta_k)n})$ functions. Here $\delta_k$ is the smallest gap between two probabilities of heads distinguishable in $k$ independent tosses with error at most $\epsilon$ under either bias. Uniformly for $3\le d\le n$, the number of degree-$d$ functions is asymptotic to $(2^{d+1}-2)\binom{n+1}{d+1}$. Almost all such functions are defined on exactly $d+1$ weights. In each fixed positive randomized cost class, a uniformly chosen function requires linearly many exact quantum queries with probability tending to one. If the prescribed exact quantum cost is $1\ll T\ll n$ and the randomized cost is $1\le r=o(\min\{T^2,\sqrt n\})$, the logarithm of the joint count is asymptotic to $T\log(n/T)$.}
\ifjournalstyle\maketitle\fi

\section{Introduction}
In the Deutsch--Jozsa problem, a truth table of length $n=2^m$ is promised to be constant or balanced. The algorithm must distinguish these cases by querying entries~\cite{dj92,cemm98}. The answer depends only on the Hamming weight of the table, the number of entries equal to one. For an arbitrary input length $n$, a symmetric partial Boolean function assigns labels to selected weights. The strings at those weights form its promise. Each of the $n+1$ weights can therefore carry a zero, a one, or no label, giving $3^{n+1}$ distinct symmetric partial Boolean functions. In each query model we count functions by the minimum number of queries they require. We also count them by their minimum polynomial degree. Each fixed positive exact quantum cost occurs only polynomially often in $n$, while each fixed positive randomized cost occurs exponentially often at fixed error $0<\epsilon<1/2$.

Qiu and Zheng determined the optimal exact cost for a family of broader Deutsch--Jozsa promises~\cite{qz18} and characterized the symmetric partial functions computable with one exact query~\cite{qiu}. On general partial functions, Xu and Qiu bounded the number of one-query functions depending on a prescribed number of input bits~\cite{xuentropy}. They also found a degree-one representative for each one-query function under their equivalence~\cite{xuijfcs}. Here each assignment of labels to a promise counts separately, and we count functions at every fixed positive exact quantum query complexity.

Sensitivity gives a related classification. For a nonempty promise, sensitivity is the maximum, over promised inputs, of the number of single-bit flips leading to an oppositely labelled promised input. Butler, Sasao and Nagayama counted total symmetric functions of maximum sensitivity~\cite{butlersens}. Xu et al. gave counting bounds by sensitivity for both total and partial symmetric functions, including an exact count at sensitivity zero for partial functions~\cite{xusens}. Query cost separates functions even at sensitivity zero. For $n\ge2$, if only weights $0,n$ are specified, equal labels require no query and opposite labels require one. The sensitivity is zero in either case.

Write $Q_E$ for the least number of quantum queries needed for certainty and $R_\epsilon$ for the least number of queries sufficient for a randomized algorithm to achieve error at most $\epsilon$ on every promised input. By the one-query classification~\cite{qiu}, the number of functions with $Q_E=1$ is $\Theta(n)$. Theorem~\ref{ra:fixed} gives $\Theta(2^{2n/3})$ functions with $R_{1/3}=1$. For a $k$-query exact quantum algorithm, the acceptance probability averaged over each weight layer is a low-degree polynomial~\cite{bbcmw}. When $n\ge2k$, a symmetric sum-of-squares decomposition~\cite{blek} keeps this polynomial in $[0,1]$ on a central real interval. At each specified weight in the interior it attains one of these bounds, so its derivative vanishes there.

We use $a\ll b$ in growth conditions to mean $a/b\to0$. Logarithms are natural unless a base is shown. At weight $t$, a uniformly sampled bit is one with probability $t/n$. Independent draws with replacement thus give coin tosses with bias $t/n$. The gap $\delta_k$ in Table~\ref{tab:maincounts} is the smallest difference between two biases distinguishable in $k$ tosses with error at most $\epsilon$ under either bias.
\begin{table}[htbp]
\caption{Counts of symmetric partial Boolean functions as $n\to\infty$. The first two rows count functions up to constant factors for fixed $k\ge1$, with $0<\epsilon<1/2$ fixed in the randomized model. In the last row the ratio of the count to the displayed expression tends to one uniformly for $3\le d\le n$.}\label{tab:maincounts}
\small\centering
\begin{tabular*}{\linewidth}{@{}p{.24\linewidth}@{\extracolsep{\fill}}p{.33\linewidth}p{.36\linewidth}@{}}\toprule
Prescribed value & Number of functions & What determines the count\\\midrule
$Q_E=k$\newline Theorem~\ref{ov:qcount} & $\Theta_k(n^k)$ & Fix the first $k$ specified weights in $[k,n-k]$ and their labels, or all such weights and labels if there are fewer. The number of completions is bounded in terms of $k$ alone.\\\addlinespace
$R_\epsilon=k$\newline Theorem~\ref{ra:fixed} & $\Theta_{k,\epsilon}(2^{(1-\delta_k)n})$ & A change of label forces a gap. Outside it, many weights can be included or omitted.\\\addlinespace
Free real degree $d$\newline Theorem~\ref{dg:all} & $\bigl(2^{d+1}-2\bigr)\binom{n+1}{d+1}$ to relative error $o(1)$ & Choose $d+1$ weights and give them a nonconstant labelling. These choices account for almost the whole class.\\\bottomrule
\end{tabular*}
\end{table}

On the full weight grid, Paturi characterized the $1/3$-approximate degree of a nonconstant symmetric function, up to constant factors, by the adjacent change of label whose midpoint lies closest to $n/2$~\cite{paturi92}. For exact representation, von zur Gathen and Roche studied how far the degree can fall below $n$~\cite{gathenroche}. Vanishing binomial sums describe when the coefficient of the highest possible power is zero~\cite{r20bcb}. In their paper, Sun and Wang give the lower bound $n-O(n^{1/3}(\log n)^{2/3})$ for nonconstant total symmetric functions~\cite{sunwang}.

For variable support, the third row of Table~\ref{tab:maincounts} gives the leading term at each free real degree in the stated range. Any labelling on $d+1$ weights has an interpolant of degree at most $d$. By Theorem~\ref{dg:all}, lower-degree nonconstant labellings on $d+1$ weights and degree-$d$ functions defined on more weights are both negligible relative to the table entry. Both estimates hold uniformly for $3\le d\le n$.

If a function is chosen uniformly at a fixed positive randomized cost, the chance that $Q_E$ falls below a sufficiently small positive multiple of $n$ decays exponentially (Corollary~\ref{ra:linearquantum}). If $1\ll T\ll n$ and $1\le r=o(\min\{T^2,\sqrt n\})$, the class with $Q_E=T$ and $R_\epsilon=r$ has the same logarithmic leading term $T\log(n/T)$ as the class with $Q_E=T$ alone. Prescribing free degree $T$ within the joint class preserves this leading term (Theorem~\ref{rd:doublebudget}). Finding the fraction of functions of quantum cost $T$ with randomized cost $r$ requires a sharper estimate. For $R_{1/3}=2$, Corollary~\ref{rtwo:all} gives an exact count for each $n\ge2$, also broken down by real degree.

The relative counts of lower-degree labellings on sampled supports and functions of the target degree on larger supports bound the sampling error. Choose $d+1$ weights uniformly, then choose a nonconstant labelling uniformly. Uniformly for $3\le d\le n$, the difference between the probability of any event under this rule and under uniform sampling at degree $d$ tends to zero. When $1\ll d\ll\sqrt n$, the two normalized query costs of a uniformly sampled degree-$d$ function have a joint limit determined by the largest score among pairs of oppositely labelled weights (Corollary~\ref{sj:jointlaw}). For oppositely labelled weights $t,t+g$ with $g>0$, the score is $\sqrt{u(1-u)}/g$, where $u=t/n$.

\section{Weight representations and complexity measures}\label{ov:model}

For a bit string $x$, its Hamming weight $|x|=\sum_{i=1}^n x_i$ counts its ones. A \emph{weight layer} consists of all strings of one weight. The \emph{weight representation} of a symmetric partial Boolean function is the word $w=(w_0,\ldots,w_n)$ over $\{0,1,*\}$. The entries $0,1$ label specified layers, and $*$ marks a layer outside the promise. This bijection lets us index functions by their words. For example, $(0,*,1,*,*)$ labels the all-zero string zero and each of the six weight-two strings one. This four-bit function has two specified weights and seven promised inputs. Reversing the word or complementing its labels gives a different function whenever the resulting word differs. There are $2^{n+1}$ total symmetric Boolean functions and $3^{n+1}$ symmetric partial Boolean functions when stars are allowed, including the empty promise. Throughout, $n\ge1$ and $0<\epsilon<1/2$.

A \emph{query} reads one input bit, or applies the quantum oracle once. The query complexity (or query cost) of a function in a given model is $k$ if it can be computed with at most $k$ queries and cannot be computed with fewer. Its randomized query complexity $R_\epsilon$ is the least nonnegative integer $k$ for which a randomized decision tree makes at most $k$ queries on every run and errs with probability at most $\epsilon$ on each promised input. The error probability is over the tree's random choices. Statements about typical functions use a uniform draw from the stated class.

The exact quantum query complexity $Q_E$ is the fewest oracle calls needed to return the correct label with certainty on every promised input. We allow input-independent operations and a final measurement. A phase query acts by $O_x|i,z\rangle=(-1)^{x_i}|i,z\rangle$, where $i$ is the address and $z$ is a workspace register. On a superposition of addresses, one oracle call negates each amplitude whose addressed bit is one. At the dummy address $i=0$ we set $x_0=0$. A bit query maps $|i,b\rangle$ to $|i,b\oplus x_i\rangle$ for $1\le i\le n$. On the Hadamard target states $|+\rangle=(|0\rangle+|1\rangle)/\sqrt2$ and $|-\rangle=(|0\rangle-|1\rangle)/\sqrt2$, it acts by phases $1$ and $(-1)^{x_i}$, respectively. To simulate the bit query, route the $|+\rangle$ branch to the dummy address, apply one phase query, then undo the routing. Conversely, to simulate a phase query, keep the original address and use an auxiliary target in $|-\rangle$ for real addresses. On the dummy branch, query a fixed real address with target $|+\rangle$. Uncomputing the auxiliaries leaves phase $(-1)^{x_i}$ on each branch, including phase one on the dummy.

Throughout the counting results, real degree means \emph{free real degree}. It is the least degree $d$ of a real polynomial matching the labels at the specified weights. The polynomial is unrestricted at omitted weights. Equivalently, the free real degree is the least total degree of a multilinear polynomial in the input bits matching the function on all promised inputs. Averaging a monomial on $j$ distinct bits over weight $t$ gives $\binom tj/\binom nj$, a polynomial of the same degree. Conversely, substituting $t=\sum_i x_i$ and using $x_i^2=x_i$ gives a multilinear polynomial of no greater degree on the cube.
\begin{table}[htbp]
\caption{Notation used across sections.}\label{tab:notation}
\small\centering
\begin{tabular*}{\linewidth}{@{}p{.26\linewidth}@{\extracolsep{\fill}}p{.70\linewidth}@{}}\toprule
Symbol & Meaning\\\midrule
$n,\ N=n+1$ & Number of input bits and number of available weights $0,\ldots,n$.\\
$w=(w_0,\ldots,w_n)$ & Weight representation. Each $w_t$ is $0$, $1$ or $*$ (weight $t$ is omitted).\\
$Q_E,\ R_\epsilon$ & Minimum query counts for an exact quantum algorithm and a randomized algorithm with worst-case error at most $\epsilon$, respectively.\\
$d,\ d(w)$ & Free real degree of the function.\\
$\delta_k$ & Smallest difference between two coin biases distinguishable in $k$ tosses with error at most $\epsilon$ for either bias (Theorem~\ref{ra:fixed}).\\
$L_d$ & Number of ways to label exactly $d+1$ weights with both labels present: $(2^{d+1}-2)\binom{N}{d+1}$.\\
$q_n(a,b)$ & Exact quantum query cost of distinguishing weights $a$ and $b$, where $0\le a<b\le n$ (Section~\ref{nav:pairs}).\\\bottomrule
\end{tabular*}
\end{table}

Consider the four-bit example above. The line $t/2$ fits both labels, so the function has free degree one. The bounded-degree convention~\cite{qiu,pyyfine} requires values in $[0,1]$ on the entire cube. The unique interpolating line takes value two at weight four, while $t(4-t)/4$ fits the labels and stays in range. The bounded degree is therefore two.

The same function has $R_{1/3}=Q_E=1$. For a randomized test, read one uniformly chosen bit. If the observed bit is one, report one. If it is zero, report one with probability $1/3$. The error is $1/3$ at weight zero and $(1/2)(2/3)=1/3$ at weight two. For an exact quantum test, start with equal amplitudes on the four real addresses. At weight zero a phase query preserves this state. At weight two it flips two signs, so the inner product with the starting state is $(1+1-1-1)/4=0$ for every such input. Measure whether the state lies in the starting direction, accepting the orthogonal outcome. Both labels occur, so zero queries cannot suffice in either model. The acceptance probability is one minus the squared overlap with the starting state, giving the quadratic above. In general, a $Q_E$-query algorithm has symmetrized acceptance probability of degree at most $2Q_E$~\cite{bbcmw}.

We assign degree zero to the function with empty domain and to constant functions. A zero-query test has the same output distribution on every input and cannot achieve error below $1/2$ on both labels. Thus all three measures have the same zero class. It consists of the empty-domain function and two constant labellings for each nonempty choice of layers, giving $1+2(2^{n+1}-1)$ functions.

The \emph{support} is the set of all specified weights. Write $M$ for its size when counting specified weights. A \emph{change of label} occurs when two successive specified weights have different labels. Omitted weights may lie between them. In the four-bit example, weights zero and two form a change of label across an omitted layer.

Table~\ref{tab:notation} collects the notation used across sections, and Table~\ref{tab:terminology} distinguishes terms that affect the counts. We write $(a)_j=a(a-1)\cdots(a-j+1)$ with $(a)_0=1$, and $\mathbf1_E$ for the indicator of an event or condition $E$. Total variation distance $\|p-q\|_{\rm TV}$ is the largest absolute difference between the probabilities that $p$ and $q$ assign to the same event. On a discrete space it equals $\tfrac12\sum_x|p(x)-q(x)|$.
\begin{table}[htbp]
\caption{Terms whose distinction affects the counts.}\label{tab:terminology}
\small\centering
\begin{tabular*}{\linewidth}{@{}p{.32\linewidth}@{\extracolsep{\fill}}p{.64\linewidth}@{}}\toprule
Terms & Meaning in this paper\\\midrule
Promise and support & The promise consists of input strings. The support is the set of their weights, including weights labelled zero.\\
Query budget and query cost & A budget of $k$ allows up to $k$ queries. Cost $k$ means that $k$ suffice and fewer cannot suffice.\\
Oppositely labelled weights and a change of label & The two labels differ in both cases. A change also requires that no other specified weight lies between the two weights.\\
Free and bounded degree & Both require the polynomial to match the specified labels. Bounded degree also requires its values to lie in $[0,1]$ at every integer weight $0,\ldots,n$.\\\bottomrule
\end{tabular*}
\end{table}

\section{Counting functions by exact quantum query complexity}\label{ax:section}
Fix an integer $k\ge1$. A function of exact quantum cost $k$ can have more than $k$ specified weights. Nevertheless, once the first $k$ specified weights in $[k,n-k]$ and their labels are known (or all of them if there are fewer), the number of completions is bounded in terms of $k$ alone. The choice of these positions accounts for the exponent $k$ in the count.

\subsection{Counts at fixed and growing exact quantum cost}
The classification of Qiu and Zheng~\cite{qiu} gives the linear count when $k=1$. In odd dimension every two-layer function of exact quantum cost one includes an endpoint and has a gap of at least $\lceil n/2\rceil$.

\begin{theorem}[Counts at prescribed exact quantum cost]\leavevmode\label{ov:qcount}%
The number of symmetric partial Boolean functions with $Q_E=k$ is $\Theta_k(n^k)$ for each fixed integer $k\ge1$. If $k\to\infty$ and $k=o(n)$, the logarithm of this count is $k\log(n/k)+O(k)$.
\end{theorem}

For a growing sublinear cost, the additive error in the logarithm is of lower order than the main term.

\begin{proof}\phantomsection\label{pr:ov:qcount}
For the lower bound, label weight zero by zero, label any $k$ weights in $[\lceil n/2\rceil,n]$ by one, and omit the rest. A single query distinguishes zero from any one selected weight $r$. Prepare a state with total squared amplitude $n/(2r)$ distributed uniformly over the real addresses and the remainder on the dummy. At weight zero the state is unchanged. At weight $r$ its inner product with the initial state is $1-2r[n/(2r)]/n=0$. Apply the $k$ tests on separate registers and accept if any test detects a change. Weight zero passes all the tests unchanged, while each selected weight is detected by its own test.

To rule out fewer queries for each constructed function, fix the final state on the all-zero input. Its inner product with the final state on a variable input is linear in the latter state's amplitudes, so its degree is at most the query count~\cite{bbcmw}. Opposite outputs require orthogonal final states. Averaging the real part of this inner product on each weight gives a polynomial equal to one at zero and zero at each of the $k$ selected weights. It therefore has degree at least $k$. The construction gives $\binom{\lfloor n/2\rfloor+1}{k}$ distinct functions of exact quantum cost $k$.

To bound the number of possible interior positions, let $P(t)$ be the acceptance probability averaged over inputs of weight $t$. The acceptance probability is a sum of squared moduli of amplitudes, so $\deg P\le2k$. The values of a polynomial at integer weights may all lie in $[0,1]$ without forcing a horizontal tangent at a zero. For example, $(t-1)(t-2)/6$ belongs to $[0,1]$ at all five integers from zero to four, yet it crosses zero at $t=1$ and becomes negative before $t=2$.

Blekherman's decomposition gives the missing interval bound. When $2k\le n$, each symmetrized acceptance or rejection probability is a sum of weighted squares~\cite[Theorem~2.3]{blek} of the form
\begin{equation}\label{qc:harmonic}
   \sum_{j=0}^{k}(t)_j(n-t)_j\sum_v b_{jv}(t)^2,
   \qquad \deg b_{jv}\le k-j .
\end{equation}
For each $j$, the inner sum runs over a finite family of real polynomials $b_{jv}$. Each probability is a sum of squares of the real and imaginary parts of final amplitudes. Layer averaging symmetrizes this sum. The amplitudes have degree at most $k$, so the decomposition applies. Both falling factorials in each prefactor of~\eqref{qc:harmonic} are nonnegative on $[k-1,n-k+1]$. At $k=2$, for instance, the two nonconstant prefactors are $t(n-t)$ and $t(t-1)(n-t)(n-t-1)$, both nonnegative on $[1,n-1]$. 

The acceptance and rejection polynomials sum to one at $n+1$ integers, hence identically because their degrees are at most $2k\le n$. Their nonnegativity throughout the interval gives $0\le P\le1$ there. Each specified integer $a\in[k,n-k]$ is in the interior of this interval. Since $P(a)$ is zero or one, it is an extremal value, and $P'(a)=0$.

Hermite interpolation uses these values and slopes together. At $k=2$, suppose the interior weights $a<b$ carry labels zero and one. In the coordinate $u=(t-a)/(b-a)$, the cubic $3u^2-2u^3$ has the required endpoint values and zero slopes. A quartic fitting the same data differs from this cubic by a multiple of $(t-a)^2(t-b)^2$. Thus the four conditions leave one free coefficient. The same double-root argument works with $k$ specified weights.

Assume $n\ge4k$, as holds eventually in both regimes. The $2k$ weights outside $[k,n-k]$ admit at most $3^{2k}$ assignments. If fewer than $k$ interior weights are specified, their positions and labels determine the function on this interval.

Otherwise let $a_1<\cdots<a_k$ be the first $k$ specified interior weights. Their prescribed values and zero derivatives determine a unique polynomial $P_0$ of degree less than $2k$. Every possible acceptance polynomial has the form $P=P_0+\lambda V$, where $V(t)=\prod_{i=1}^k(t-a_i)^2$, because the difference has a double root at each $a_i$.

If there is another specified interior weight, we must bound the choices for its position independently of $n$. Let $y>a_k$ be the first such weight and $b$ its label. Since $V(y)\ne0$, the value condition gives $\lambda=(b-P_0(y))/V(y)$. Substituting this into the slope condition $P_0'(y)+\lambda V'(y)=0$ gives
$P_0'(y)V(y)-(P_0(y)-b)V'(y)=0$.
For either label the left side has degree at most $4k-2$. If it vanishes identically, $(P_0-b)/V$ is constant by the quotient rule. Since $\deg P_0<\deg V$, that constant is zero and $P_0=b$. The value at $y$ would then force $\lambda=0$ and $P=b$, impossible for a function taking both values. In every admissible case there are consequently only $O(k)$ choices of $(y,b)$.

Once the extra weight and its label fix $P$, all later specified interior weights are roots of $P'$. Among these at most $2k-1$ roots, only those where $P$ is zero or one are eligible, and $P$ fixes their labels. After the first $k$ positions and their labels are known, we may therefore stop, or choose one of $O(k)$ pairs $(y,b)$ and a subset of at most $2k-1$ roots. Including the labels on the first positions and the exterior assignments gives at most $2^k(1+O(k)2^{2k-1})3^{2k}=\exp(O(k))$ completions for each initial choice of positions. The same bound covers fewer than $k$ such positions.

The total count is at most $\exp(O(k))\sum_{i=0}^k\binom{n+1}{i}$. For fixed $k$ this upper bound matches the order given by the lower-bound construction. For $k=o(n)$ the binomial sum is bounded by $(k+1)(e(n+1)/k)^k$, while $\binom{\lfloor n/2\rfloor+1}{k}\ge(n/(2k))^k$. Taking logarithms gives the stated uniform error.
\end{proof}

\subsection{Implementing bounded polynomials with phase queries}
The joint counting argument will construct functions by prescribing an acceptance polynomial. The next lemma converts its degree bound into a query budget if the polynomial maps the entire unit interval into itself.

\begin{lemma}[Realizing a bounded polynomial]\leavevmode\label{qa:real}%
Let $n\ge1$ and $T\ge0$ be integers. Suppose a real polynomial $p$ has degree at most $2T$ and satisfies $0\le p(u)\le1$ for every real $u\in[0,1]$. In the phase-oracle model with a dummy address (Section~\ref{ov:model}), there is an algorithm that uses at most $T$ queries and, on every input $x\in\{0,1\}^n$, accepts with probability exactly $p(|x|/n)$.
\end{lemma}

For example, after one phase query on the uniform address state, the inner product with the starting state is $1-2u$ at normalized weight $u=|x|/n$. Projecting onto the orthogonal complement of the starting state therefore realizes $p(u)=1-(1-2u)^2=4u(1-u)$. To realize an arbitrary $p$, the proof factors both $p$ and $1-p$ and builds their amplitudes one query at a time.

\begin{proof}\phantomsection\label{pr:qa:real}
Put $z=e^{i\theta}$ and $t=(1-\cos\theta)/2$. Real Fej\'er--Riesz factorization gives $p(t)=|F(z)|^2$ and $1-p(t)=|G(z)|^2$, with real coefficients and degrees at most $2T$. Indeed, multiply each nonzero nonnegative Laurent polynomial of order $h$ by $z^h$. The resulting real reciprocal polynomial has roots paired by inversion and conjugation. Take all roots inside the disk. Nonnegativity gives each unit-circle root even multiplicity, so take half its copies. The selected roots are closed under conjugation. The factor and its reversal recover the polynomial up to a positive scale fixed on the circle. For constants, including zero, take the nonnegative square root.

Following the factorization idea of generalized quantum signal processing~\cite{motlagh}, form the unit vector $v(z)=2^{-1/2}\big(z^{-T}F(z),z^T F(1/z),z^{-T}G(z),z^T G(1/z)\big)$. Its first two coordinates are conjugates with total squared norm $p(t)$. Shifting by $T$ keeps all powers between $-T$ and $T$. If $v(z)=\sum_{j=-h}^h v_j z^j$ has order $h$, its real coefficients satisfy $v_{-j}=Bv_j$, where $B$ swaps the coordinates within each pair.

To remove the highest and lowest powers of $z$, suppose $h>0$. Every nonconstant Fourier coefficient of $\|v\|^2=1$ vanishes. At frequency $2h$ this gives $v_{-h}^{\mathsf T}v_h=0$. At frequency $2h-1$ the two contributions $v_{-h}^{\mathsf T}v_{h-1}$ and $v_{1-h}^{\mathsf T}v_h$ agree by the symmetry under $B$. Their sum vanishes, so each vanishes, also when $h=1$.

Let $D_+,D_-$ project onto the spans of $v_h,v_{-h}$, respectively, and put $D_0=I-D_+-D_-$. These control projections depend only on the polynomial coefficients and require no input queries. Multiplication by $z^{-1}D_++zD_-+D_0$ shifts the two projected components in opposite directions. The coefficients of $z^{h+1}$ and $z^h$ become $D_-v_h=0$ and $D_-v_{h-1}+D_0v_h=0$ by the orthogonality just proved. Since $B$ exchanges $D_+$ and $D_-$, the coefficients at $-h-1$ and $-h$ vanish as well, and the coefficient symmetry persists. The multiplier is unitary on the circle, so the norm remains one. Each step removes at least one positive and one negative frequency. Reversing at most $T$ steps recovers the original vector from a constant one.

To implement these inverse steps at weight $|x|$, put $t=|x|/n$ and let $|s\rangle=n^{-1/2}\sum_{j=1}^{n}|j\rangle$ be the uniform state on the real addresses. Initialize the coefficient register in the constant vector obtained above. Write $R_s=2|s\rangle\langle s|-I$ for the reflection about $|s\rangle$. The Grover unitary $V=R_sO_x$ and its inverse $V^\dagger=O_xR_s$ differ only in the placement of this reflection. When both bit values occur, take the uniform states on the addresses of zeros and of ones. On their span the Grover unitary has eigenphases $\pm\theta$, determinant one and trace $2(1-2t)$. At the endpoints, the starting state has eigenvalue $1$ or $-1$.

Each inverse step requires the three controlled operations $V,V^\dagger,I$ on the ranges of $D_+,D_-,D_0$. All three branches share a single call to $O_x$. On the $D_-$ branch apply the reflection before the call, and on the $D_+$ branch apply it afterwards. On the $D_0$ branch route the address to the dummy using an auxiliary register and undo the routing after the call. On the Grover eigenspace for $z$, this step applies $zD_++z^{-1}D_-+D_0$, reversing the reduction. After all steps the first two coordinates have total squared norm $|F(z)|^2=p(t)$, since $F(1/z)=\overline{F(z)}$ on the unit circle. Accept on the first two coordinates. The eigenphase components are orthogonal and have the same acceptance probability, hence so does the starting address state $|s\rangle$.
\end{proof}

\subsection{Functions rejecting only weight zero}\label{nav:zero-family}
For $1\le d\le n$, take a $d$-subset $S\subseteq\{1,\ldots,n\}$. The function $f_S$ rejects weight zero, accepts each weight in $S$, and is undefined at every other weight. Its free real degree is $d$. Indeed, for any interpolant $p$, the polynomial $p-1$ has $d$ distinct roots and is nonzero at zero, while $d+1$ values admit an interpolant of degree at most $d$. For the joint count, we determine the randomized cost of these $\binom nd$ functions and bound their exact quantum cost using polynomials.

Fix $0<\epsilon<1/2$, let $m=\min S$ and put $L=\log((1-\epsilon)/\epsilon)$. After a random permutation of the coordinates, a tree with query cap $0\le j\le n$ can be padded to sample $j$ distinct bits uniformly without replacement. At weight $t>0$, the all-zero record has probability $h_j(t)=\binom{n-t}{j}/\binom nj$. Among the accepting weights this probability is largest at $t=m$. At weight zero only the all-zero record is possible. If the rule accepts that record with probability $p$, its maximum error is at least $\max\{p,(1-p)h_j(m)\}$. Accept every other record and set $p=h_j(m)/(1+h_j(m))$. This equalizes the errors at zero and the smallest accepting weight and attains the bound. Hence $R_\epsilon(f_S)\le j$ exactly when $h_j(m)\le e^{-L}$.

Let $a_0=n+1$. For $1\le j\le n$ let $a_j$ be the least positive integer $m$ passing this test. The randomized cost is $j$ precisely when $a_j\le m<a_{j-1}$, so

\begin{equation}\label{ax:W}
 N_j:=\#\{S:|S|=d,\ R_\epsilon(f_S)=j\}
   =\binom{n-a_j+1}d-\binom{n-a_{j-1}+1}d .
\end{equation}
The convention $\binom{q}{d}=0$ for integers $0\le q<d$ also covers empty classes. For example, take $n=6$, $d=2$ and $\epsilon=1/3$. One read suffices when $m\ge3$, since $h_1(m)=(6-m)/6$ must be at most $1/2$. Two reads suffice at $m=2$, since $h_2(2)=6/15$. They do not suffice at $m=1$, where $h_2(1)=10/15$. Exactly the four sets $S=\{2,r\}$, $r=3,4,5,6$, need two reads, agreeing with \eqref{ax:W} since $\binom52-\binom42=4$.

To locate the integer thresholds, first bound their real counterparts. On $0\le s<n-j+1$, let $G_j(s)=-\sum_{i=0}^{j-1}\log(1-s/(n-i))$. It is strictly increasing from zero to infinity, with derivative $G'_j(s)=\sum_{i=0}^{j-1}(n-i-s)^{-1}$. The unique solution of $G_j(s_j)=L$ gives $a_j=\lceil s_j\rceil$, even when $h_j$ vanishes at this integer. Bounding $n-i$ between $n-j+1$ and $n$ gives
\begin{equation}\label{qc:threshold}
  (n-j+1)(1-e^{-L/j})\le s_j\le n(1-e^{-L/j}) .
\end{equation}
The two bounds differ by at most $L$ since $j(1-e^{-L/j})\le L$.

The minimum of $S$ determines the randomized cost. For the exact quantum cost the relevant polynomial has a root at every member of $S$.

\begin{lemma}[Polynomial bounds for functions rejecting only weight zero]\leavevmode\label{ax:signal}%
Let $S\subseteq\{1,\ldots,n\}$ have size $d\ge1$, let $T\ge0$ be an integer and set $F(t)=\prod_{r\in S}(1-t/r)$. Consider real polynomials $P$ of degree at most $T$ satisfying $P(0)=1$ and $P(r)=0$ for all $r\in S$.
\begin{enumerate}
\item If $Q_E(f_S)\le T$, such a $P$ exists with $|P(j)|\le1$ for every integer $0\le j\le n$. It has the form $P=FU$, with $U(0)=1$ and $\deg U\le T-d$.
\item If such a $P$ satisfies $|P(t)|\le1$ for every real $t\in[0,n]$, then $Q_E(f_S)\le T$.
\end{enumerate}
\end{lemma}

\begin{proof}\phantomsection\label{pr:ax:signal}
Store measurement records in extra registers to preserve purity, and pad to $T$ queries with the dummy address. A query multiplies an addressed amplitude by $1-2x_i$. Between queries, input-independent unitaries form linear combinations of amplitudes. Hence each final amplitude is multilinear of degree at most $T$ on the cube~\cite{bbcmw}. Take the real part of the inner product between the final states on the given and all-zero inputs. This polynomial equals one at zero and vanishes on inputs of weight in $S$, since exactness requires orthogonal states for opposite outputs. Cauchy--Schwarz bounds its modulus by one. On a layer of weight $t$, the average of a square-free monomial in $j$ bits is $(t)_j/(n)_j$. The layer average is a univariate polynomial of degree at most $T$ and modulus at most one at every integer weight. Its $d$ distinct roots in $S$ permit division by $F$. The quotient has degree at most $T-d$ and value one at zero because $F(0)=1$.

For the second implication we use the bound on the whole real interval. It makes $p(u)=1-P(nu)^2$ an admissible acceptance probability throughout $[0,1]$, with value zero at $u=0$ and one at $u=r/n$ for every $r\in S$. Its degree is at most $2T$, so Lemma~\ref{qa:real} applies.
\end{proof}

For example, let $n=6$, $S=\{1,6\}$ and $T=2$. The two roots and the value at zero force $P(t)=(1-t)(1-t/6)$. Its values at the seven integer weights are $1,0,-2/3,-1,-1,-2/3,0$, but $P(7/2)=-25/24$. Thus this candidate meets the grid bound in the first part, while the second part cannot certify a two-query algorithm for this promise.

\subsection{Joint counts by quantum and randomized query cost}

At $T=d=|S|$ the only candidate in Lemma~\ref{ax:signal} is $P=F$. Thus choosing $S$ so that $|F|\le1$ on $[0,n]$ gives exact quantum cost $T$, while the minimum of $S$ fixes the randomized cost. This family gives the lower bound. For the upper bound, we use the stationary points forced by successive weights with the same label.

\begin{theorem}[Joint counts by quantum and randomized query cost]\leavevmode\label{rd:doublebudget}%
Fix $0<\epsilon<1/2$ and take positive integer sequences $T=T(n)$, $r=r(n)$ with $T\to\infty$, $T=o(n)$ and $r=o(T^2)$. Let $\mathcal F$ be either of the following classes of symmetric partial Boolean functions.
\begin{enumerate}
\item For a prescribed randomized cost, take $\mathcal F=\{w:Q_E(w)=T,\ R_\epsilon(w)=r\}$ and assume $r=o(\sqrt n)$.
\item For a randomized query budget, take $\mathcal F=\{w:Q_E(w)=T,\ 1\le R_\epsilon(w)\le r\}$ and assume $r=o(n)$.
\end{enumerate}
Writing $d(w)$ for the free real degree, the counts satisfy
\begin{equation}\label{qc:jointlog}
 \log|\mathcal F|\sim\log\#\{w\in\mathcal F:d(w)=T\}
     \sim T\log(n/T).
\end{equation}
For $w$ drawn uniformly from the chosen class, let $X$ be any one of $d(w)$, the size of its support, or the number of its support weights in $[T,n-T]$. For each fixed $0<\delta<1$,
\begin{equation}\label{qc:jointconcentration}
 \Pr\bigl(|X/T-1|>\delta\bigr)
 \le \exp[-(\delta-o(1))T\log(n/T)] .
\end{equation}
\end{theorem}
In either class, prescribing free degree equal to quantum cost preserves the leading logarithmic count in~\eqref{qc:jointlog}. The three ratios to quantum cost each tend to one in probability. Under $r=o(T^2)$, label changes are negligible relative to $T$. For randomized cost exactly $r$, the construction uses distinct integer thresholds for $r-1$ and $r$. The assumption $r=o(\sqrt n)$ guarantees their separation.
\begin{proof}\phantomsection\label{pr:rd:doublebudget}
Let $M$ count the specified weights, $I$ those in $[T,n-T]$, and $s$ the number of label changes between successive specified weights. Every label change requires the randomized acceptance probability to move by at least $1-2\epsilon$. We bound how far that probability can move in total as the weight runs from zero to $n$. Permute the coordinates uniformly and pad every run to $r$ distinct draws. At a given weight all sample orders with the same number of ones have the same probability, so the sample sum suffices to describe the acceptance probability.

Let $H_t$ denote the sample-sum distribution at weight $t$. Couple weights $t$ and $t+1$ by flipping one zero and sampling the same coordinates in both inputs. When the sample misses this bit (probability $1-r/n$), the sums agree and their common contribution cancels. When it hits the bit, the sums are $Y,Y+1$, where $Y$ is hypergeometric with $t$ ones among the $n-1$ other bits and $r-1$ draws. The hypergeometric mass function is unimodal, so the sum of the absolute differences of its successive masses is twice its peak. Thus the distance between $Y$ and $Y+1$ is $\max_j\Pr(Y=j)$, and $\|H_{t+1}-H_t\|_{\rm TV}=(r/n)\max_j\Pr(Y=j)$.

For $r\ge2$, $r=o(n)$ and $u=t/(n-1)$, the variance of $Y$ is $(r-1)u(1-u)(n-r)/(n-2)$. Log-concavity bounds its largest point probability by $O((1+\operatorname{Var}Y)^{-1/2})$~\cite{r26bmm}. At $t=0,n-1$, each distance is $r/n$. For $1\le t\le n-2$, the sum of $[u(1-u)]^{-1/2}$ is $O(n)$, giving $\sum_{t=0}^{n-1}\|H_{t+1}-H_t\|_{\rm TV}=O(\sqrt r)$. For $r=1$ the distances sum to one. The test takes values in $[0,1]$, so its expectations at consecutive weights differ by at most $\|H_{t+1}-H_t\|_{\rm TV}$. Summing over the disjoint gaps where labels change gives $(1-2\epsilon)s=O(\sqrt r)$, so $s=o(T)$.

We bound the central and total support sizes by counting derivative roots. For $T=o(n)$, the interval bound following~\eqref{qc:harmonic} in the proof of Theorem~\ref{ov:qcount} bounds the acceptance polynomial between zero and one on $[T-1,n-T+1]$. Both labels occur since $R_\epsilon\ge1$. Each of the $I$ specified weights in $[T,n-T]$ is thus an interior extremum. Rolle's theorem gives a further derivative root inside each of the $M-1-s$ gaps with equal endpoint labels. These open gaps are disjoint and contain no specified weight. The derivative of the nonconstant acceptance polynomial therefore has at least $I+(M-1-s)$ distinct roots and degree at most $2T-1$. This gives $I+M\le2T+s$. The derivative of a minimum-degree interpolant also has a root in each equal-label gap, so $M-1-s\le d-1$. Interpolation on the $M$ specified weights gives the other inequality in $M-s\le d\le M-1$.

It now suffices to count functions with few central weights. A value of $d$, $M$ or $I$ below $(1-\delta)T$ gives $I\le(1-\delta)T+s$ by $I\le M\le d+s$. If $d$ or $M$ exceeds $(1+\delta)T$, then so does the support size by interpolation. Thus $I+M\le2T+s$ gives the same bound. Finally $2I\le I+M\le2T+s$ rules out $I>(1+\delta)T$ eventually. Thus every exceptional function has $I\le(1-\delta)T+O_\epsilon(\sqrt r)$.

There are at most $3^{2T}\sum_{i\le b}2^i\binom{n+1}{i}$ functions with at most $b$ specified central weights. At the cutoff for exceptional functions, the logarithm of this bound is at most $(1-\delta+o(1))T\log(n/T)$. For the total count, $2I\le2T+s$ and the uniform bound $s=O_\epsilon(\sqrt r)$ allow a common cutoff $b=T+O_\epsilon(\sqrt r)$. The resulting logarithmic upper bound is $(1+o(1))T\log(n/T)$.

For the matching lower bound take $|S|=T$ in the family $f_S$. Any choice of $S$ already gives $d=T$ and, by the necessary polynomial condition in Lemma~\ref{ax:signal}, $Q_E\ge T$. For the class with prescribed randomized cost, choose the minimum to obtain $R_\epsilon=r$ and the remaining elements so that the product $\prod_{w\in S}(1-t/w)$ has modulus at most one on $[0,n]$. The sufficient condition in the same lemma will then give $Q_E\le T$.

Rounding~\eqref{qc:threshold} gives $a_r=n(1-e^{-L/r})+O_\epsilon(1)$ uniformly, since $r(1-e^{-L/r})\le L$. For $r\ge2$ the magnitude of the derivative of $1-e^{-L/x}$ on $[r-1,r]$ is at least $Le^{-L}/r^2$. Thus $a_{r-1}-a_r\ge Le^{-L}n/r^2-O_\epsilon(1)>1$ eventually under the stronger growth condition when randomized cost is prescribed exactly. Taking $a=a_r$ as the minimum gives exactly $r$ randomized queries, including $r=1$.

The factor from a small minimum $a$ can exceed one in modulus far from zero. We reserve a few roots to control it and choose the rest from a fixed interval of large weights. Normalize the weights by $n$, write $u=a/n$ and fix $b_0\in(\max(3/4,1-e^{-L}),1)$. Every root $w\in[b_0,1]$ gives $|1-y/w|\le e^{-y}$ on $[0,1]$. For $y\le w$ use $1-y/w\le1-y$, and for $y\ge w$ use $y/w-1\le1/3<e^{-1}$.

If $u\ge1/16$, fix only $u$, whose factor is bounded by $e^{16y}$, and put $J=1$. Otherwise fix the $J=\lfloor1/(2\sqrt u)\rfloor$ roots $u,4u,\ldots,J^2u$. For $0\le y\le1/8$, Euler's product for $\sin(\pi z)/(\pi z)$ at $z=\sqrt{y/u}$ bounds the fixed product in modulus by $\exp(Cy/\sqrt u)$. The infinite product has modulus at most one. Dividing by its tail recovers the fixed product. The reciprocal of the tail is at most $\exp[(2y/u)\sum_{i>J}i^{-2}]$ since $y/(i^2u)<1/2$ for $i>J$. By integral comparison, the sum of inverse squares is at most the reciprocal of the cutoff. Each prescribed root $i^2u$ is an actual normalized weight, since $nu=a$ is an integer. For $1/8\le y\le1$, an imaginary argument in the same product bounds the first $J$ factors in modulus by $\prod_{i\ge1}(1+y/(i^2u))\le e^{\pi/\sqrt u}$.

Concavity gives $1-e^{-L/r}\ge(1-e^{-L})/r$. Together with $u=1-e^{-L/r}+O_\epsilon(1/n)$ and $r=o(n)$, this bound gives $u^{-1/2}=O_\epsilon(\sqrt r)=o(T)$, so $J=o(T)$. Choosing $T-J$ distinct roots above $b_0$ supplies a factor at most $e^{-(T-J)y}$. The fixed factors contribute at most $\exp(O(y/\sqrt u))$ throughout $[0,1]$, since $1\le8y$ on the range $y\ge1/8$. Since $u^{-1/2}=o(T)$ and $T-J\sim T$, decay from the free roots offsets any growth from the fixed roots. The full product thus has modulus at most one. Also, $u\le1-e^{-L}+O_\epsilon(1/n)<b_0$ eventually. In the case with additional fixed roots, each is at most $J^2u\le1/4$. The free and fixed roots are therefore distinct.

All these functions now have $d=Q_E=T$ and $R_\epsilon=r$. Choosing $T-J$ free roots among $(1-b_0)n+O(1)$ weights gives a family of logarithmic size $(1+o(1))T\log(n/T)$. This supplies the lower bound even in the subclass $d=T$. Subtracting this lower bound on the logarithm of $|\mathcal F|$ from the upper bound on the logarithm of the exceptional count gives~\eqref{qc:jointconcentration}. For $1\le R_\epsilon\le r$, the construction with minimum $a_1$ already gives the required lower bound. The upper-bound argument uses only $r=o(n)$ and $r=o(T^2)$.
\end{proof}

\section{Counting functions by randomized query complexity}\label{ra:sec}

Fix $0<\epsilon<1/2$. At weight $t$, the unknown input contains $t$ ones in $n$ positions. In a uniform sample of $k\le n$ distinct coordinates, all information about $t$ is contained in the number $j$ of observed ones. For example, the records $01$ and $10$ are equally likely at every weight compatible with them, so distinguishing their order cannot improve a test.

To see why this also covers adaptive trees, apply a uniform permutation to the addresses, reuse the answers to repeated queries and pad early-stopping runs to $k$ distinct reads. Permuting coordinates preserves the promise and the error bound. Fix the original tree's random choices, leaving the address permutation random. Each new address reveals a uniformly chosen remaining bit, even if earlier answers determine the address. Conditional on $j$, all orders are equally likely. Averaging over these orders and the random choices gives the acceptance probability $u_j$, independent of $t$. The acceptance probability at weight $t$ is therefore
\begin{equation}\label{ra:sampling}
 B_k(t)=\sum_{j=0}^k u_j\,
 \frac{\binom tj\binom{n-t}{k-j}}{\binom nk}.
\end{equation}
Conversely, sampling $k$ coordinates and responding with these probabilities realizes \eqref{ra:sampling}. We use $\binom mi=0$ for $m\ge0$ and $i\notin[0,m]$.

\subsection{Growth at a fixed positive randomized cost}\label{nav:randomwidth}

Take $k=1$, $\epsilon=1/3$ and assume $3$ divides $n$. Return one uniformly chosen bit. At weight $n/3$ its error is $1/3$ for label zero, and at weight $2n/3$ the same error holds for label one. Fix these labels and omit all intervening weights. Each of the $n/3$ weights below $n/3$ may be omitted or labelled zero, and each of the $n/3$ above $2n/3$ may be omitted or labelled one. This gives $2^{2n/3}$ functions of randomized cost one. Without the divisibility assumption, rounding the boundary weights outwards changes the exponent by at most two.

For general $k$, suppose that the input has a fraction $x$ of ones and sample $k$ coordinates independently with replacement. The number of observed ones is binomial. Sampling distinct coordinates instead gives the hypergeometric distribution in~\eqref{ra:sampling}. For fixed $k$, the coupled samples agree unless the independent draws repeat an address (probability at most $\binom{k}{2}/n$).

The rule that accepts on seeing at least $m$ ones has acceptance probability $P_m(x)=\Pr(\mathrm{Bin}(k,x)\ge m)$. For $1\le m\le k$, $x_m(v)=P_m^{-1}(v)$ is the input fraction of ones at which the rule accepts with probability $v$. Thus $x_m(\epsilon)$ is the largest fraction that can be labelled zero and $x_m(1-\epsilon)$ the smallest that can be labelled one. Their minimum separation over $m$ is $\delta_k$. For a minimizing threshold, round $nx_m(\epsilon)$ down and $nx_m(1-\epsilon)$ up. Once these two weights are labelled, each of the $(1-\delta_k)n+O(1)$ exterior weights may independently be omitted or receive its side's label.

\begin{theorem}[Counts at fixed positive randomized query complexity]\leavevmode\label{ra:fixed}%
For the binomial-tail inverse $x_m(v)$ defined above, let $\delta_k=\min_{1\le m\le k}[x_m(1-\epsilon)-x_m(\epsilon)]$. For fixed $k\ge1$ and $0<\epsilon<1/2$, the number of symmetric partial functions with $R_\epsilon=k$, including total functions, is $\Theta_{k,\epsilon}(2^{(1-\delta_k)n})$ as $n\to\infty$. The first two widths are $\delta_1=1-2\epsilon$ and $\delta_2=\sqrt{1-\epsilon}-\sqrt\epsilon$. The formula $\delta_k=(1-\epsilon)^{1/k}-\epsilon^{1/k}$ holds for every $k\ge1$ when $1/4\le\epsilon<1/2$. For each fixed $0<\epsilon<1/2$, the same formula holds for every $k\ge k_0(\epsilon)$.
\end{theorem}
The main counting issue is the position of a change of label. With endpoints fixed, a gap $j$ weights wider than the shortest feasible gap leaves $2^{-j}$ times as many exterior choices. The proof bounds the number of possible starting weights by $O_{k,\epsilon}(j+1)$, so summing over positions preserves the constant-factor estimate.
\begin{proof}\phantomsection\label{pr:ra:fixed}
Consider the binomial experiment with error bound $0<v<1/2$. We seek the closest fractions $x<y$ that can be assigned opposite labels. For $0<x<y<1$, the likelihood ratio in favour of $y$ increases with the sample sum $S$. At $x=0$ or $y=1$ the same ordering holds by cross-multiplication. Among rules with the same probability of wrongly accepting at $x$, a threshold rule therefore maximizes the probability of accepting at $y$~\cite{r12np}. It outputs one for $S>m$, zero for $S<m$, and one with probability $\alpha$ on the tie $S=m$. Write its acceptance probability as $p_\alpha(z)=\Pr_z(S>m)+\alpha\Pr_z(S=m)$.

For $0<m<k$ the shortest gap for a fixed tie probability $\alpha$ has endpoints $l,r$ given by $p_\alpha(l)=v$ and $p_\alpha(r)=1-v$. Implicit differentiation gives $l'=-A(l)$ and $r'=-A(r)$, where $A(z)=[\alpha m/z+(1-\alpha)(k-m)/(1-z)]^{-1}$. Thus the derivative of $r-l$ has the sign of
\begin{equation}\label{ra:tiewidthsign}
 \frac{(1-\alpha)(k-m)}{(1-l)(1-r)}-\frac{\alpha m}{lr}.
\end{equation}
As $\alpha$ increases, both quantiles decrease. The first term decreases and the magnitude of the second increases, so the derivative changes sign once, from positive to negative. The minimum occurs at a deterministic endpoint, $\alpha=0$ or $1$. At each endpoint the width has a strictly positive inward one-sided derivative.

At $m=0$, $p_\alpha(z)=1-(1-\alpha)(1-z)^k$ attains both target levels only for $0\le\alpha\le v$. The width is $[(1-v)^{1/k}-v^{1/k}](1-\alpha)^{-1/k}$, minimized at $\alpha=0$. At $m=k$, $p_\alpha(z)=\alpha z^k$ attains both levels only for $1-v\le\alpha\le1$. The width is then $[(1-v)^{1/k}-v^{1/k}]\alpha^{-1/k}$, minimized at $\alpha=1$. These two cases already cover $k=1$, with minimum width $1-2v$.

To pass from budgets to exact costs, we show that the minimum distinguishable gap decreases strictly with $k$. From $k+1$ reads, discard one uniformly at random and apply a minimizing $k$-read rule $S\ge m$. With $k+1$ observations, the rule accepts above $m$. At $m$ it accepts with probability $(k+1-m)/(k+1)$. This probability is strictly between zero and one, with $0<m<k+1$. By the sign change in~\eqref{ra:tiewidthsign}, at least one deterministic response to the tie gives a strictly narrower gap. Thus $\delta_{k+1}<\delta_k$.

To count nonconstant functions with $R_\epsilon\le k$ from below, draw $k$ addresses with replacement and apply a minimizing threshold at error $\epsilon$. Reuse earlier answers at repeated addresses to stay within $k$ queries. Label $a=\lfloor nx_m(\epsilon)\rfloor$ zero and $b=\lceil nx_m(1-\epsilon)\rceil$ one. Monotonicity gives $P_m(a/n)\le\epsilon$ and $P_m(b/n)\ge1-\epsilon$, including equality at an integer boundary. Omit the weights strictly between $a$ and $b$. Each exterior weight can be omitted or given the label on its side, yielding $2^{a+n-b}$ nonconstant functions. Since $b-a=n\delta_k+O(1)$, this is $\Omega(2^{(1-\delta_k)n})$.

For the upper bound, use the preceding coupling of $k$ draws. Independent sampling repeats an address with probability at most $\binom{k}{2}/n$. A test with error at most $\epsilon$ on distinct draws therefore has error at most $v=\epsilon+\binom{k}{2}/n$ on independent draws. For large $n$ we have $v<1/2$. Denote the minimum width at this error by $\delta_k(v)$. Since $P_m'$ is positive at both interior quantiles, the finitely many widths and their minimum are locally Lipschitz in $v$. It follows that $n\delta_k(v)=n\delta_k+O_{k,\epsilon}(1)$. An adjacent pair of oppositely labelled specified weights must be distinguishable in this binomial experiment. Its gap is therefore at least $n\delta_k-O_{k,\epsilon}(1)$.

Functions with at least two changes contribute exponentially less than the lower-bound family. The polynomial in \eqref{ra:sampling} has degree at most $k$ and crosses $1/2$ between consecutive opposite labels, so at most $k$ changes occur. There are $O_k(n^{2k})$ choices of their endpoints. The first two open gaps are disjoint and together exclude at least $2n\delta_k-O(1)$ weights. Once the first label is chosen, each other weight may be omitted or receive its run's label, giving at most $O_k(n^{2k})2^{(1-2\delta_k)n+O(1)}$ functions.

It remains to count functions with one change of label. Complement the outputs if needed. Let $x$ and $y$ be the normalized weights of the last zero and first one, and put $s=n(y-x)-n\delta_k(v)\ge0$. For a feasible threshold at the enlarged error $v$, let $p_\alpha(l)=v$ and $p_\alpha(r)=1-v$. Since $x\le l$ and $y\ge r$, the excess width is a sum of three nonnegative terms,
\begin{equation}\label{ra:excesswidth}
 \frac{s}{n}=(l-x)+\bigl(r-l-\delta_k(v)\bigr)+(y-r).
\end{equation}
The first and last terms are the unused space on the two sides of the test interval $[l,r]$. The middle term is the test's excess width above the minimum. A nearly shortest gap thus forces both a nearly optimal test and a starting weight close to its left quantile.

To quantify this, fix the tie level $m$ and write $W_v(\alpha)=r-l$. At $v=\epsilon$ there are only finitely many minimizing endpoints $\alpha_*$. The one-sided derivatives above and the explicit formulas for $m=0,k$ give $W_v(\alpha)-W_v(\alpha_*)\ge c|\alpha-\alpha_*|$ near each minimizer, with a common $c>0$ for $v$ close to $\epsilon$. Outside these neighbourhoods, compactness leaves a positive gap above the minimum, uniformly for such $v$. For small $s/n$, \eqref{ra:excesswidth} excludes distant thresholds. Near a minimizer, $W_v(\alpha_*)\ge\delta_k(v)$ and the local bound give $|\alpha-\alpha_*|\le s/(cn)$ for some minimizing endpoint at error $\epsilon$.

The two quantiles vary smoothly with $(v,\alpha)$ near this endpoint because the binomial tail has positive derivative at each quantile. Since $v-\epsilon=O_k(1/n)$, the left quantile differs by $O_{k,\epsilon}((s+1)/n)$ from its value at the minimizing endpoint for the original error. The first term in \eqref{ra:excesswidth} gives $l-x\le s/n$. Consequently the integer starting weight $nx$ lies within $O_{k,\epsilon}(s+1)$ of one of finitely many locations.

For $j\le s<j+1$, there is at most one integer gap $n(y-x)$. The preceding bound leaves $O_{k,\epsilon}(j+1)$ starting positions. For $s/n$ bounded away from zero the same bound follows from the $n+1$ available positions. For each choice of endpoints, the $n-n(y-x)$ exterior weights can each be omitted or given the label on their side. Allowing both orientations, the count in this range is $O_{k,\epsilon}((j+1)2^{n-n\delta_k(v)-j})$. The sum over $j$ converges after the common exponential factor is removed, since $\sum_{j\ge0}(j+1)2^{-j}=4$.

The lower and upper bounds now match for $1\le R_\epsilon\le k$. To count functions of exact cost $k$, remove those computable with fewer queries. For $k\ge2$, the count for $1\le R_\epsilon\le k-1$ is $O_{k,\epsilon}(2^{(1-\delta_{k-1})n})$, exponentially smaller because $\delta_{k-1}>\delta_k$. For $k=1$ the nonconstant functions counted already require a query.

It remains to identify the minimizing threshold in the two stated ranges. The outer thresholds $m=1$ and $m=k$ both have width $D=(1-\epsilon)^{1/k}-\epsilon^{1/k}$. For each interior threshold the acceptance probability rises by $1-2\epsilon$ across the gap, so its width is at least $(1-2\epsilon)/\max P'_m$. Here $\max P'_m=kp_i$, where $i=m-1$, $q=k-1$ and $p_i=\binom qi(i/q)^i(1-i/q)^{q-i}$. For consecutive interior indices, $p_{i+1}/p_i=(1+1/i)^i/(1+1/(q-i-1))^{q-i-1}$. The numerator and denominator come from the same increasing sequence, so the ratio is at most one on the left half. Symmetry then gives $p_i\le p_1\le4/9$ for $k\ge4$. Thus each interior width times $k/(1-2\epsilon)$ is at least $9/4$.

For $\epsilon\ge1/4$, $kD/(1-2\epsilon)$ is the mean of $x^{1/k-1}$ on $[\epsilon,1-\epsilon]$. Averaging the majorant $1/x$ over $[1/4,3/4]$ gives the upper bound $2\log3<9/4$. Both intervals are centred at $1/2$, and enlarging a centred interval increases the mean of the convex function $1/x$. For $k=3$ the interior bound is $2$, while the chord bound for the mean of $x^{-2/3}$ is $(4^{2/3}+(4/3)^{2/3})/2<2$. For $k\le2$ there is no interior threshold.

For fixed $\epsilon$, symmetry reduces the interior thresholds to $2\le m\le(k+1)/2$. Stirling's bounds give $p_{m-1}=O(m^{-1/2})$, so $k$ times an interior width exceeds $L+1$ for large $m$, where $L=\log((1-\epsilon)/\epsilon)$. The scaled endpoint width tends to this logarithm, leaving finitely many indices $m$ independent of $k$.

For each of these fixed $m\ge2$, $1-P_m(y/k)\to S_m(y)=e^{-y}\sum_{i=0}^{m-1}y^i/i!$. The scaled quantiles converge to $a,b$ satisfying $S_m(a)=1-\epsilon$, $S_m(b)=\epsilon$. Integrating $-(\log S_m)'<1$ from $a$ to $b$ gives $b-a>L=\lim_{k\to\infty}kD$. Each of the finitely many remaining interior thresholds is therefore eventually wider than an endpoint threshold.
\end{proof}
The count at fixed randomized cost gives the denominator for the proportion requiring few exact quantum queries.
\begin{corollary}[Typical exact quantum cost at fixed randomized cost]\leavevmode\label{ra:linearquantum}%
Fix a positive integer $k$ and $0<\epsilon<1/2$. For each sufficiently large $n$, choose $f$ uniformly among the functions on $n$ bits with $R_\epsilon(f)=k$. There are constants $c,\eta>0$ depending only on $k,\epsilon$ such that
$\Pr_f(Q_E(f)\le cn)\le e^{-\eta n}$.
Thus $cn<Q_E(f)\le n$ except on a set of probability at most $e^{-\eta n}$.
\end{corollary}

\begin{proof}\phantomsection\label{pr:ra:linearquantum}
Write $\mathcal C_k$ for the class $R_\epsilon=k$. To bound the exceptional fraction, first count labelled promises in $\mathcal C_k$ with $Q_E\le cn$. The symmetrized acceptance polynomial $P$ of an optimal exact algorithm has degree at most $2cn$~\cite{bbcmw}. Each zero-labelled weight is a root of $P$, and each one-labelled weight is a root of $P-1$. Both polynomials are nonzero because both labels occur. Thus the two root sets together contain at most $4cn$ specified weights.

There are at most $\sum_{i=0}^{\lfloor4cn\rfloor}2^i\binom{n+1}{i}$ such functions. For a summand of index $i$, choose $i$ weights and then choose their labels. For $0<4c<1/2$ the logarithm of this upper bound is at most $n[H(4c)+4c\log 2+o(1)]$, where $H(x)=-x\log x-(1-x)\log(1-x)$.

For the whole class, Theorem~\ref{ra:fixed} gives $\log|\mathcal C_k|=(1-\delta_k)n\log2+O_{k,\epsilon}(1)$. Its linear coefficient is at least $2\epsilon\log 2$ because $\delta_k\le\delta_1=1-2\epsilon$. Choosing $c$ small makes $H(4c)+4c\log2$ strictly smaller than this coefficient. The positive gap between these two growth rates gives the exponent $\eta$ after division by $|\mathcal C_k|$. Reading all $n$ bits gives $Q_E\le n$ for every function.
\end{proof}

\subsection{Exact counts at two randomized queries and error one third}\label{rtwo:sec}
At error $1/3$, functions of randomized query complexity two can be counted for each $n$. A two-query randomized test for such a function has a symmetrized acceptance polynomial of degree at most two, which crosses $1/2$ at each change of label. Positive cost requires both labels, so the label changes once or twice. With two changes, the middle block has the opposite label to both outer blocks. The Deutsch--Jozsa promise is an even-dimensional example, with zeros at weights $0,n$ and a one at the middle weight.

Qiu and Zheng classified symmetric promises with one exact quantum query or bounded polynomial degree one or two~\cite{qiu}. For odd $n\ge3$, distinguishing either central weight, or their union, from the endpoints requires exactly two quantum queries~\cite{qzpre}. Podder, Yao and Ye studied the four-layer case with a reflected central pair~\cite{pyyfine}. At randomized cost two, Theorem~\ref{rtwo:turns} counts the functions with two changes of label and compares their free real degree with their exact quantum cost. Corollary~\ref{rtwo:all} adds the functions with one change. The two proofs appear in Appendix~\ref{app:twoproofs} in the same order.

\begin{theorem}[Two changes at error one third]\leavevmode\label{rtwo:turns}%
For $n\ge2$ let $W=\{t\in\{0,\ldots,n\}: |2t-n|\le\sqrt n\}$. Consider functions with $R_{1/3}=2$ and exactly two changes of label on their specified weights.

\textup{(i)} The class has $2(2^{|W|}-1)+e_n$ members. Each nonempty $S\subseteq W$ gives a function rejecting at $0,n$ and accepting on $S$. These functions and their output complements account for $2(2^{|W|}-1)$ members of the class. For $n\ge6$, every exceptional function, up to input reflection and output complementation, rejects at $1,n$, may also reject at $0$, and accepts on a nonempty subset of $[n/2,n/2+1]\cap\mathbb Z$. All other weights are unspecified in both descriptions. The number $e_n$ of exceptions is zero for $n\le5$, $24$ for even $n\ge6$, and $8$ for odd $n\ge7$.

\textup{(ii)} For a function constructed from $S$ in part (i), or its output complement, put $s=|S|$ and $D_S=\prod_{t\in S}(n-t)/t$. Its free degree and exact quantum cost are
\begin{equation}\label{rtwo:endpointcost}
 d=s+\mathbf1_{\{(-1)^sD_S\ne1\}},\qquad
 Q_E=s+\mathbf1_{\{D_S\ne1\}}.
\end{equation}
They differ exactly when $s$ is odd and $D_S=1$. This never occurs in odd dimension. In even dimension at least $2^{(|W|-1)/2}$ choices of $S$ give $Q_E=d-1$.

\textup{(iii)} As $n\to\infty$, the number of functions in the full two-change class with $Q_E=d=k$ is $(1+o(1))2\binom{|W|}{k-1}$ uniformly for $3\le k\le |W|$. At $k=2$ the count is $(3+o(1))|W|$. For a uniform draw from the full class, $\Pr(Q_E=d)\to1$. Each of $(Q_E-|W|/2)/(\sqrt{|W|}/2)$ and $(d-|W|/2)/(\sqrt{|W|}/2)$ converges in distribution to a standard normal random variable.
\end{theorem}

Free real degree and exact quantum cost can differ even when the only rejecting weights are $0,n$. To see this in the four-bit example of Section~\ref{ov:model}, also label weight four zero. Here $S=\{2\}$ and $D_S=1$, so \eqref{rtwo:endpointcost} gives $d=2$ and $Q_E=1$. A uniform phase query preserves the starting direction at weights zero and four and gives an orthogonal state at weight two. For a randomized test, read two distinct bits and accept if they differ. This accepts with probability $4/6$ at weight two and zero at the endpoints. One read gives an affine acceptance probability, whose value at the midpoint cannot exceed both endpoint values. Thus $R_{1/3}=2$.

For the rest of this subsection and Appendix~\ref{app:twoproofs}, use the same central set $W$ as in Theorem~\ref{rtwo:turns}. Up to output complementation, a single change runs from zero to one. The last zero $a$ and first one $b$ determine how many randomized queries suffice. For $n\ge2$, $q=1,2$ and $0\le a<n$, let $b_q(a)$ be the least weight $b>a$ distinguishable from $a$ with at most $q$ randomized queries at error $1/3$, or $n+1$ if there is none.

For one draw, the acceptance probability is affine in the normalized weight. The line through $(a/n,1/3)$ and $(b/n,2/3)$ stays in $[0,1]$ on the unit interval precisely when $b\ge2a$ and $2b\ge n+a$. Any feasible line has at least this slope, takes a value at most $1/3$ at $a/n$ and at least $2/3$ at $b/n$. At the left endpoint it lies no higher than the line through the two boundary points, and at the right endpoint it lies no lower. The first weight $b$ satisfying the endpoint constraints is $b_1(a)$, using $n+1$ if no such weight exists.

For $b_2(a)$, write $P_i(t)=\binom ti\binom{n-t}{2-i}/\binom n2$ for the two-draw point probabilities in~\eqref{ra:sampling} and consider observations $i=2,1,0$ in that order. Accept outcomes with certainty until including the next would raise the acceptance probability at $a$ above $1/3$. Randomize at that outcome to reach $1/3$ and reject all lower outcomes. Higher observations impossible at $a$ are always accepted. Comparing consecutive hypergeometric ratios (by cross-multiplication at a support endpoint) shows that $P_i(b)/P_i(a)$ is nondecreasing in $i$. By the Neyman--Pearson lemma~\cite{r12np}, this rule maximizes power at every $b>a$, that is, acceptance probability there subject to the error bound at the lower weight. If such a weight exists, $b_2(a)$ is the least $b>a$ at which the power is at least $2/3$. Sampling the same coordinates after flipping one zero to one cannot decrease the count, so the threshold rule works at every lower zero and higher one.

Let $A_i$ count the nonempty sets $S\subseteq W$ of size $i$ for which $D_S=1$. Set $A_i=0$ outside $1\le i\le |W|$. Equation~\eqref{rtwo:endpointcost} assigns degree $|S|+1$ except when $|S|$ is even and $D_S=1$, in which case the degree is $|S|$. The next two formulas each count three disjoint groups in order. These are the functions with one change, those constructed from $S\subseteq W$ and their output complements, and the exceptions counted by $e_n$.

\begin{corollary}[Counts by real degree for functions of randomized query complexity two]\leavevmode\label{rtwo:all}%
For every $n\ge2$ the number of functions with $R_{1/3}=2$ is
\begin{equation}\label{rtwo:totalcount}
 2\sum_{a=0}^{n-1}\!\left(2^{n+a-b_2(a)+1}-2^{n+a-b_1(a)+1}\right)+2(2^{|W|}-1)+e_n .
\end{equation}
For an integer $j\ge1$, the number with both $R_{1/3}=2$ and free degree $j$ is
\begin{equation}\label{rtwo:degreecount}
\begin{aligned}
 &2\sum_{a=0}^{n-1}\!\left[\binom{n+a-b_2(a)+1}{j}-\binom{n+a-b_1(a)+1}{j}\right]\\
 &\hspace{2em}+2\left[\mathbf1_{\{j\ge2\}}\binom{|W|}{j-1}+\mathbf1_{\{2\mid j\}} A_j-\mathbf1_{\{2\mid j-1\}} A_{j-1}\right]+e_{n,j}.
\end{aligned}
\end{equation}
Here $\binom hj=0$ when $j>h\ge0$. The correction $e_{n,j}$ counts the exceptions at degree $j$. It vanishes for $n\le5$ and for $j\notin\{2,3,4\}$. For even $n\ge6$, $(e_{n,2},e_{n,3},e_{n,4})=(12,8,4)$. For odd $n\ge7$, the corresponding triple is $(4,4,0)$.
For $1\le i\le |W|$, the counts $A_i$ are the coefficients of $z^i\prod_p X_p^0$ in
\begin{equation}\label{rtwo:balancedproduct}
 \prod_{t\in W}\left(1+z\prod_{\substack{p\le n\\p\ \mathrm{prime}}} X_p^{v_p(n-t)-v_p(t)}\right),
\end{equation}
where $v_p(m)$ is the exponent of the prime $p$ in $m$. At $n=1$ the class $R_{1/3}=2$ is empty.
\end{corollary}

For example, at $n=4$ the threshold lists for $a=0,1,2,3$ are $b_1=(2,3,4,5)$ and $b_2=(1,3,4,4)$. For $a=1$ the optimal two-query rule has $(u_0,u_1,u_2)=(0,2/3,1)$. Its acceptance probability is $11/18$ at $b=2$ and $5/6$ at $b=3$. The one-change contribution by degree is $4z(1+z)^3$. In $W=\{1,2,3\}$ the nonempty sets with $D_S=1$ are $\{2\},\{1,3\},W$, giving $8z^2+4z^3+2z^4$ for the two-change contribution. The degree generating polynomial for the full class is $4z+20z^2+16z^3+6z^4$. Evaluating at $z=1$ gives $46$ functions.

\section{Counting functions of prescribed free real degree}\label{dg:section}

Any labelling on $m$ specified weights, with $m\ge1$, has an interpolating polynomial of degree at most $m-1$. The support size is bounded above because a nonconstant degree-$d$ polynomial takes zero and one at most $d$ times each. Thus the support size $m$ of a function of degree $d$ satisfies $d+1\le m\le\min(2d,n+1)$ for $d\ge1$. For $d\ge3$, counting the nonconstant labellings on $d+1$ weights gives the correct leading term, uniformly even when $d$ grows with $n$.

Write $N=n+1$, let $A$ and $Z$ be the sets of weights labelled one and zero, and put $D=A\cup Z$. For nonempty $D$, interpolate the labels with the unique polynomial of degree below $|D|$. Its degree is the free real degree of the function (Section~\ref{ov:model}). Each labelling of $D$ counts separately.

\subsection{Counting and sampling functions of degree three or higher}\label{dg:leadingsec}
The sampling rule chooses $d+1$ weights uniformly and then chooses a nonconstant labelling uniformly on those weights.

\begin{theorem}[Uniform enumeration by free real degree]\leavevmode\label{dg:all}%
Let $L_d=(2^{d+1}-2)\binom{N}{d+1}$. As $n\to\infty$, uniformly for $3\le d\le n$, the number of symmetric partial Boolean functions of free real degree $d$ is $(1+o(1))L_d$. The uniform law on this class is within $o(1)$ in total variation of the law obtained by choosing $d+1$ weights uniformly and giving them a uniform nonconstant labelling. For fixed $d\ge3$ the absolute counting error is $O_{d,\eta}(N^{d+1/2+\eta})$ for every $\eta>0$.
\end{theorem}

There are $\binom{N}{d+1}$ choices of support and $2^{d+1}-2$ nonconstant labellings on each. To recover the degree-$d$ count from $L_d$, subtract the nonconstant labellings on $d+1$ weights whose interpolant has degree below the prescribed value, then add the degree-$d$ functions with larger supports. The corrections can cancel in the count, but their sum counts functions in exactly one class. Bounding each correction relative to $L_d$ therefore controls the total variation distance, the largest absolute difference between the probabilities assigned to any one event. At degree two, a fraction tending to $1/7$ of the functions have four specified weights (Proposition~\ref{dg:finite}).

For a finite set $S$ put $P_S(t)=\prod_{u\in S}(t-u)$. The Lagrange basis polynomial at $a\in D$ is $P_D(t)/[(t-a)P_D'(a)]$. Summing these polynomials over $a\in A$ gives the interpolant. For $m=|D|$, its coefficient of $t^{m-1}$ is $\sum_{a\in A}1/P_D'(a)$. For example, on $D=\{0,1,2,3\}$ the reciprocal derivatives in increasing weight order are $(-1,3,-3,1)/6$. Labelling only weights $1$ and $2$ by one cancels the cubic coefficient. The quadratic $t(3-t)/2$ gives precisely these labels.

For $m\ge2$, each $a$ contributes $(a-\sum_{u\in D}u)/P_D'(a)$ to the coefficient of the next lower power. On the full grid, $1/P_D'(a)$ is a constant multiple of $\omega_a=(-1)^a\binom na$. The coefficients of the two highest possible powers vanish together precisely when $\sum_{a\in A}\omega_a=\sum_{a\in A}a\omega_a=0$. Von zur Gathen and Roche give the full sequence of moment conditions for total functions~\cite[Theorem~2.2]{gathenroche}. When only some weights are specified, the denominators depend on the chosen support, and we must count the choices of $D$ as well as its labellings.

\begin{proof}\phantomsection\label{pr:dg:all}
Write $C_j$ for the number of degree-$d$ functions with $j$ specified weights. The loss is $L_d-C_{d+1}$, and the gain is $\sum_{j\ge d+2}C_j$. We estimate the loss on $d+1$ weights and bound the total gain using the count on $d+2$ weights. For $d\le N^{3/5}$ we count completions of partially revealed supports at these two sizes. For larger $d$ we estimate the probability of leading-coefficient cancellation at each size.

Retaining any $d+2$ weights preserves degree $d$, since the interpolating polynomial is already uniquely determined by its values on any $d+1$ of them. It can take Boolean values at no more than $2d$ weights in total, leaving at most $d-2$ possible extensions of the restriction. Counting deletions of $r$ weights from a labelled support of size $d+2+r$ gives
$C_{d+2+r}\binom{d+2+r}{r}\le C_{d+2}\binom{d-2}{r}$.
After division by the number of deletions from the larger support, each factor in the binomial ratio for $0\le r\le d-2$ is at most $1-(r+4)/(d+2+r)$. The ratio is therefore at most $\exp[-r^2/(2d+3)]$. Summing over $r$ and comparing with the Gaussian integral bounds the entire gain by $O(\sqrt d\,C_{d+2})$.

Suppose first that $d\le N^{3/5}$. To estimate the loss, draw a uniform nonconstant labelling on a uniform support of size $d+1$. Hide a uniformly chosen weight carrying the more frequent label, breaking ties between labels by a fixed rule. Both labels remain because $d+1\ge4$ and the deleted label occurred at least twice. Given the label counts and the retained labelled support, every completion has the same deletion probability. The hidden weight is therefore uniform on the $N-d$ unused positions. A drop below degree $d$ occurs exactly when the interpolant of the $d$ retained values takes the hidden label at that weight. This interpolant is nonconstant and has degree at most $d-1$, so at most $d-1$ positions work. The lost fraction is at most $(d-1)/(N-d)$.

For the gain, start with a function counted by $C_{d+2}$. Both labels occur at least twice, since each occurs at most $d$ times. Delete one accepting weight $x$ and one rejecting weight $y$, and denote the nonempty retained label sets by $A,Z$. Let $q$ be their interpolant of degree below $d$ and put $G=P_AP_Z$. Every polynomial of degree at most $d$ fitting the retained labels has the form $q+cG$ because its difference from $q$ vanishes at all $d$ retained weights.

Eliminating the common coefficient $c$ from $(q+cG)(x)=1$ and $(q+cG)(y)=0$ gives $U(y)=V(x)$, where $U=q/G$ and $V=(q-1)/G$. Factor $q=P_ZH$ and $q-1=P_AK$. The fractions $U=H/P_A$ and $V=K/P_Z$ are reduced, since $q=1$ on $A$ and $q-1=-1$ on $Z$. Thus every admissible pair $(x,y)$ is an integer point of the plane curve
\begin{equation}\label{dg:extensioncurve}
 H(y)P_Z(x)-K(x)P_A(y)=0 .
\end{equation}
Since $\deg q<d=\deg G$, the extension $q+cG$ has degree $d$ precisely when $c\ne0$. Counting all integer points on the curve therefore gives an upper bound on the number of degree-$d$ extensions.
Both products in \eqref{dg:extensioncurve} have total degree at most $\deg q\le d-1$, since $\deg H=\deg q-|Z|$ and $\deg K=\deg q-|A|$. The poles of $U$ and $V$ in separate variables show that the defining polynomial is nonzero.

The missing pair $(x,y)$ lies in the integer square $\{0,\ldots,N-1\}^2$. For $B\ge2$, Theorem~2 of Binyamini, Cluckers and Novikov~\cite{r17bcn} gives an integer-point bound. In $[-B,B]^2$, an irreducible affine plane curve of degree $e$ over $\mathbb C$ contains at most $C e^2 B^{1/e}(\log B)^\kappa$ integer points. Here $C$ and $\kappa$ are independent of the curve and of $e$. Their corresponding estimate for rational points of bounded height has exponent $2/e$.

The exponent $1/e$ gives a square-root saving when $e\ge2$. In the square of admissible weights, the line $y=x$ contains $N$ integer points, while the parabola $y=x^2$ contains exactly $\lfloor\sqrt{N-1}\rfloor+1$. We must therefore exclude all line components of \eqref{dg:extensioncurve} before applying the bound.

A vertical component $x=x_0$ would force $K(x_0)=P_Z(x_0)=0$ or $U$ constant. The first alternative contradicts coprimality, and $U$ has a pole at every member of the nonempty set $A$. The same argument excludes horizontal lines.

A line component of finite nonzero slope would give $U(\alpha t+\beta)=V(t)$ identically, with $\alpha\ne0$. Comparing poles gives $\alpha Z+\beta=A$, hence $|A|=|Z|=r=d/2\ge2$. Two distinct real poles have real images, so $\alpha,\beta$ are real. Since $P_Z$ divides $q$, we have $r\le\deg q\le2r-1$. At infinity $U$ and $V$ share the leading term $c_0t^{-s}$, with $c_0\ne0$ and $1\le s=2r-\deg q\le r$. Comparing terms gives $\alpha^s=1$, so the real slope is $1$ or $-1$.

Consider first a translation. For $\alpha=1,\beta\ne0$ the leading term of $U(t+\beta)-U(t)$ is $-sc_0\beta t^{-s-1}$, whereas $V(t)-U(t)=-1/G(t)$ has order $t^{-2r}$ and $s+1<2r$. For $\alpha=1,\beta=0$ we would have $1/G=0$.

For a reflection, $\alpha=-1$. This involution exchanges the pole sets and preserves their even-degree monic product $G$. Both sides of $q(\beta-t)=1-q(t)$ interpolate the complementary labels on the $d$ retained weights, so uniqueness gives the identity. Thus $U(\beta-t)=-V(t)$, inconsistent with $U(\beta-t)=V(t)$ since $V$ is nonzero. This excludes every line component.

We may now apply the integer-point bound with $B=N$ and sum once over the distinct irreducible components of \eqref{dg:extensioncurve} over $\mathbb C$. Their degrees $e\ge2$ sum to at most $d-1$. For $e\le\log N$ use $e^2\le e\log N$ and $N^{1/e}\le\sqrt N$. For $e>\log N$ use $N^{1/e}\le\exp(1)$ and $\sum e^2\le d^2$. Thus \hypertarget{nav:dgextensions}{each retained labelled support} has at most $T=O((d\sqrt N+d^2)(\log N)^{\kappa+1})$ possible extensions.

To turn this extension bound into a count of functions, fix the number $i$ of retained ones. Reconstruction gives $i+1$ ones and $d-i+1$ zeros. Dividing by the $(i+1)(d-i+1)$ possible deleted pairs counts each function once. Summing over the retained labelled supports and using $\binom di/[(i+1)(d-i+1)]=\binom{d+2}{i+1}/[(d+1)(d+2)]$ gives $C_{d+2}\le\binom Nd\,T\,\frac{2^{d+2}-2}{(d+1)(d+2)}$. Here the sum was enlarged from $1\le i\le d-1$ to $0\le i\le d$.

Divide this bound by $L_d$ and use $\binom Nd/\binom N{d+1}=(d+1)/(N-d)$. Including the $O(\sqrt d)$ factor for further extensions gives a relative gain of $O(T/[(N-d)\sqrt d])$. For $d\le N^{3/5}$ this is $O((\sqrt{d/N}+d^{3/2}/N)(\log N)^{\kappa+1})=o(1)$. The relative loss is $O(d/(N-d))$. For fixed $d$, multiply both relative bounds by $L_d=O_d(N^{d+1})$ and absorb the logarithms to get the stated absolute error.

Now suppose $d>N^{3/5}$. On $d+1$ nodes, a nonconstant labelling is lost exactly when the highest interpolation coefficient vanishes. On $d+2$ nodes, every degree-$d$ function again has vanishing highest coefficient, although cancellation can also give a lower degree. One cancellation bound therefore controls the loss at the first support size and the gain at the second.

Choose $(t_1,\ldots,t_m)$ uniformly among ordered tuples of distinct weights and assign independent fair labels. Set \hypertarget{nav:dgrecip}{$a_i=1/\prod_{j\ne i}(t_i-t_j)$}. Write each label as $(1+\sigma_i)/2$ with independent fair signs $\sigma_i$. The coefficient of $t^{m-1}$ then equals a fixed constant plus $\frac12\sum_i \sigma_i a_i$. For fixed nodes, the $a_i$ are nonzero rationals. Clearing their denominators preserves the largest point probability of the sign sum. Hal\'asz's anti-concentration bound relates this probability to equal magnitudes among the coefficients~\cite{r19hal}. Let $R_1$ count the ordered choices of $i,j$ and signs $\delta,\delta'\in\{-1,1\}$ for which $\delta a_i+\delta'a_j=0$. The largest point probability is $O(m^{-5/2}R_1)$~\cite[Theorem~3.3, $k=1$]{r19tv}. For $i\ne j$, a relation requires $|a_i|=|a_j|$. Thus it remains to estimate how often two reciprocal derivatives at different nodes have the same magnitude.

There are $O(m)$ relations with $i=j$. For $i\ne j$, hold all nodes except a third node $z$ fixed. Then $a_i=c_i/(t_i-z)$ and $a_j=c_j/(t_j-z)$ with $c_i,c_j$ fixed and nonzero. The two poles are distinct, so for either relative sign the numerator of $a_i\pm a_j$ is a nonzero polynomial of degree at most one in $z$. Conditionally, $z$ is uniform on the $N-m+1$ unused positions, and at most one gives a relation with that sign. For instance, in the four-node example, replace $3$ by $z>2$. The coefficients at $1$ and $2$ become $1/(z-1)$ and $-1/[2(z-2)]$. They cancel only at $z=3$. Summing the bound on relations gives $\mathbb E R_1=O(m+m^2/(N-m+1))$. Let $E_m$ be the probability that the coefficient of $t^{m-1}$ vanishes. Averaging Hal\'asz's bound over the nodes gives
\begin{equation}\label{dg:cancellation}
 E_m=O\!\left(m^{-3/2}+\frac{m^{-1/2}}{N-m+1}\right).
\end{equation}
At $m=d+1$, divide \eqref{dg:cancellation} by $1-2^{1-m}$ to bound the loss conditional on a nonconstant labelling. The resulting bound tends uniformly to zero even at $m=N$.

For the gain, first suppose $d<n$. The coefficient of $t^{d+1}$ in the interpolant on $d+2$ weights must vanish for a function of degree $d$, so $C_{d+2}\le 2^{d+2}\binom{N}{d+2}E_{d+2}$. Dividing this binomial coefficient by $\binom N{d+1}$ gives $(N-d-1)/(d+2)$. At this support size the numerator equals $N-m+1$ and cancels the denominator in the second term of \eqref{dg:cancellation}. Including further extensions gives a relative gain of $O(N/d^2+1/d)=o(1)$ uniformly in this range. At $d=n$ there are no larger supports.

The two gain estimates explain the choice $3/5$ for the cutoff exponent. Any fixed $1/2<\alpha<2/3$ works because the relative gain from the curve estimate tends uniformly to zero up to $N^\alpha$, and $N/d^2+1/d$ tends uniformly to zero above $N^\alpha$.

Finally, for \hypertarget{nav:dgfiniteTV}{nonempty finite classes} $C,D$ their uniform laws have total variation distance $1-|C\cap D|/\max(|C|,|D|)$, by summing the smaller point mass on their intersection. Here the two classes differ on exactly $\mathrm{loss}+\mathrm{gain}=o(L_d)$ functions and their sizes are asymptotic to $L_d$. The formula gives the required uniform $o(1)$ error in the probability of any event.
\end{proof}

\subsection{Exact counts for low degrees and for one accepting weight}

The low-degree cases explain why the uniform estimate starts at degree three. With a single accepting weight, an exact count is possible at every positive degree.

\begin{proposition}[Low degrees and a single accepting weight]\leavevmode\label{dg:finite}%
There are $2\binom N2$ functions of free real degree one. The number of functions of degree two is $6\binom N3+2\sum_{a\ge1}\binom{N-2a}{2}$.
The two terms count supports of size three and four, respectively, with their labels. A binomial coefficient in the sum is zero if its top entry is below two. For each $d\ge1$, the number of degree-$d$ functions with exactly one accepting weight is $N\binom{N-1}{d}$.
\end{proposition}

\begin{proof}\phantomsection\label{pr:dg:finite}
A nonconstant line takes each label at most once. Thus its support consists of two differently labelled weights, with two choices of orientation.

A degree-two function has three or four specified weights. On three weights all six nonconstant labellings have degree two, since none fits a line. On four weights a quadratic must take each label twice. Its two equal-value pairs have the same midpoint, the axis of the parabola. Conversely, either assignment of opposite labels to the inner and outer pairs comes from an affine function of squared distance from that midpoint. In increasing order, the three gaps are therefore $a,b,a$, with $a,b\ge1$. Fix $a$. A given $b$ leaves $N-2a-b$ starting positions whenever this quantity is positive. Summing over $b$ gives $\binom{N-2a}{2}$, with two labellings for each support.

For a single accepting weight $a$, the $|Z|$ prescribed zeros force degree at least $|Z|$, and $P_Z(t)/P_Z(a)$ realizes equality. Choosing $a$ and then the $d$ zeros gives the last count.
\end{proof}

The three-weight count is asymptotic to $N^3$. Dividing the four-weight count by $N^3$ gives a Riemann sum with limit $\int_0^{1/2}(1-2x)^2\,dx=1/6$. Hence the limiting fraction of degree-two functions supported on four weights is $(1/6)/(1+1/6)=1/7$. This explains the lower endpoint $d=3$ in Theorem~\ref{dg:all}.

\subsection{Prescribing the number of accepting weights}
Section~\ref{sec:jointmeasures} uses the degree-$d$ count with exactly $r$ weights labelled one. We count every support compatible with degree $d$. The sampling rule uses $d+1$ weights, the smallest support on which that degree is possible.

\begin{proposition}[Free degree with a prescribed number of accepting weights]\leavevmode\label{dg:fixedr}%
Let $1\le d\le n$, $s=d+1$, $1\le r\le s/2$, and $k=s-r$. Let $C_{d,r}$ count degree-$d$ functions with exactly $r$ accepting weights, over all support sizes. Put $B_{d,r}=\binom Ns\binom sr$. Then
\begin{equation}\label{dg:prescribedcounts}
 C_{d,r}\ge B_{d,r}\left(1-\frac{2}{k+1}\right),
 \qquad C_{d,r}=(1+o(1))B_{d,r}\quad(d\to\infty).
\end{equation}
The asymptotic is uniform in $n\ge d$ and the indicated range of $r$.

Choose $s$ weights uniformly and label a uniform $r$-subset one and its complement in the support zero. This law and the uniform law on the $C_{d,r}$ functions have total variation distance $o(1)$ in the same range. For fixed $\delta>0$ and $d\ge\delta n$, both this distance and $|C_{d,r}/B_{d,r}-1|$ are $O_\delta(1/n)$.

For functions with $\min(|A|,|Z|)=r$, either label can occur $r$ times. In the same limit their number is asymptotic to $2B_{d,r}$ for $2r<s$ and to $B_{d,r}$ for $2r=s$.
\end{proposition}

\begin{proof}\phantomsection\label{pr:dg:fixedr}
We again count the loss on the minimum support and the gain on larger supports. This time each count must be bounded relative to $B_{d,r}$, since the prescribed number of ones may be rare among unrestricted labellings.

First bound the loss, the number of labellings counted by $B_{d,r}$ whose interpolant has degree below $d$. On $s=d+1$ weights this happens exactly when the coefficient of $t^d$ vanishes. If $s=N$, the full-grid coefficient is proportional to $\sum_{a\in A}\omega_a$, where $\omega_a=(-1)^a\binom na$. Hide one uniformly chosen accepting weight. Given the other $r-1$ ones, the hidden weight is uniform on $k+1$ positions and must have one prescribed signed binomial value to cancel the revealed sum. Since $\binom na$ increases strictly to the midpoint, a signed value occurs at most twice, giving cancellation probability at most $2/(k+1)$.

If $s<N$, hide a uniformly chosen omitted weight $v$ and a uniformly chosen accepting weight $a_*$. Let $U$ be the set of all other omitted weights. All labelled partitions with these two distinguished weights are equiprobable. Write $\omega_a(U)=\omega_aP_U(a)$ and $S_i=\sum_{a\in A}a^i\omega_a(U)$. For a retained weight $a$, the reciprocal derivative $1/P_D'(a)$ equals its full-grid value times $P_U(a)(a-v)$. Thus a loss requires $S_1-vS_0=0$. Given $U,A$, the hidden $v$ is uniform on $k+1$ positions. If $S_0\ne0$, at most one position works.

If $S_0=0$, the loss condition no longer depends on $v$ and requires $S_1=0$ as well. To bound this joint probability, reveal only $U$ and $A\setminus\{a_*\}$, leaving $v$ hidden. The weight $a_*$ has $k+2$ equally likely positions. Writing $S'_i$ for the revealed sums, the two equations give $S'_0=-\omega_{a_*}(U)\ne0$ and $S'_1=-a_*\omega_{a_*}(U)$. Only $a_*=S'_1/S'_0$ works. Splitting according to $S_0=0$ thus bounds the lost fraction by $1/(k+1)+1/(k+2)\le2/(k+1)$, proving the finite lower bound in~\eqref{dg:prescribedcounts}. These surviving labellings already have degree $d$ on exactly $s$ weights. Since $k\ge(d+1)/2$, the relative loss tends uniformly to zero as $d\to\infty$.

For the gain, a function with $s+e$ specified weights, $e\ge1$, has $r$ ones and $k+e$ zeros. Deleting $e$ zeros retains all $r$ ones and leaves $s=d+1$ values, which determine the original polynomial. It has at most $d-k=r-1$ further zeros available for extension. Each enlarged function can be recovered in $\binom{k+e}e$ ways, one for each deleted subset. For this support size, the gain relative to $B_{d,r}$ is at most $\binom{r-1}e/\binom{k+e}e$. If $r=1$, no extension is possible, in agreement with Proposition~\ref{dg:finite}. For $r=o(d)$ the sum is $O(r/k)$ because its $e$th term is at most $((r-1)/(k+1))^e$.

To prove uniformity, consider any sequence with $d\to\infty$ and pass to a subsequence on which $r/d$ has a limit. The geometric bound settles the case of a zero limit. For a positive limit, $r\asymp d$. To reduce the count to functions with one extra specified weight, repeat the deletion argument retaining $s+1$ weights. The total gain is then at most $O(\sqrt d)$ times the count on $s+1$ weights, since $\binom{r-2}e/\binom{k+1+e}e\le\exp[-e^2/(2d+3)]$.

For $d\le N^{3/5}$ delete one weight of each label from the $s+1$ weights. The \hyperlink{nav:dgextensions}{curve estimate $T$} in the proof of Theorem~\ref{dg:all} bounds the number of two-weight extensions of each set of $d$ retained labelled weights. Dividing by the number of possible deletions per function gives at most $T\binom Nd\binom d{r-1}/[r(k+1)]$ functions, or $T/[(N-d)(k+1)]$ relative to $B_{d,r}$. Even with the $O(\sqrt d)$ factor this tends uniformly to zero.

For $d>N^{3/5}$ and $s<N$, randomly order $m=s+1$ chosen weights. Pair the first $2r$ weights. For each pair, a fair coin selects the accepting weight. The remaining positions are rejecting, so exactly $r$ positions accept. By symmetry this gives a uniform $r$-subset of the unordered weights. Write $a_i$ for the \hyperlink{nav:dgrecip}{reciprocal derivative} at the $i$th weight. The $i$th pair contributes either $a_{2i-1}$ or $a_{2i}$ to the leading coefficient. Subtracting the midpoint of the two values gives a fair sign times $(a_{2i-1}-a_{2i})/2$. Conditional on the ordered weights, these signs are independent.

To apply the Hal\'asz bound we need a positive fraction of the paired differences to be nonzero. In sorted order the reciprocal derivatives alternate in sign. Before each of the first $\lfloor r/4\rfloor$ pairs, fewer than $r/2$ of the $m\ge2r$ positions have been used. Both signs therefore occupy a fixed positive fraction of the remaining positions. Opposite signs give a nonzero difference with conditional probability at least some $p_0>0$. By coupling, the number of nonzero differences among these pairs stochastically dominates $\operatorname{Bin}(\lfloor r/4\rfloor,p_0)$. Thus at least $cr$ differences are nonzero except with probability $e^{-c'r}$, for absolute $c,c'>0$.

Apply the order-one Hal\'asz bound to these nonzero differences. There are $O(r)$ trivial signed relations. For a nontrivial relation between distinct pairs, hide one weight outside their four indices (possible for large $r$) and expose the rest. As in the proof of \eqref{dg:cancellation}, each reciprocal derivative has a simple pole in the hidden weight. The four indices give distinct poles with nonzero residues, and the numerator of their signed sum has degree at most three. Each relation thus holds with probability at most $3/(N-m+1)$. Even including zero differences, the expected number of such relations is $O(r^2/(N-m+1))$. With at least $cr$ nonzero differences, the prefactor in the Hal\'asz bound is $O(r^{-5/2})$. Adding the probability of the complementary event bounds the probability of a degree drop by $O(r^{-3/2}+r^{-1/2}/(N-m+1))$.

On $m=s+1=d+2$ weights, exact degree $d$ requires the coefficient of $t^{d+1}$ to vanish. Its cancellation probability therefore gives an upper bound on the fraction to count. Across all supports, there are $(N-s)/(k+1)$ times as many labellings on $m$ specified weights with $r$ ones as on $s$ specified weights with $r$ ones. Multiply by this ratio and by $O(\sqrt d)$ for further extensions. The factor $N-s=N-m+1$ cancels the denominator in the second term of the probability bound. Since $r\asymp d$, the relative gain is $O(N/d^2+1/d)=o(1)$. At $s=N$ no larger support exists.

For $d\ge\delta n$ the lost fraction is $O_\delta(1/n)$. The gain bound has this order for $r\ge s/3$. For $r<s/3$, a degree-$d$ function with $s+1$ specified weights requires a drop below the maximum possible interpolation degree. The probability of this drop is at most $2/(k+2)$ by the loss estimate at size $s+1$. The population ratio $(N-s)/(k+1)$ is $O_\delta(1)$ and, relative to this first larger support class, further extensions sum geometrically because $r/(k+1)<1/2$. This again gives $O_\delta(1/n)$.

For odd $n\ge3$, take $d=n$ and $r=2$. The full-grid coefficients at $a$ and $n-a$ have the same magnitude and opposite signs, and no other pairs cancel. Exactly $(n+1)/2$ of the $\binom{n+1}{2}$ choices are lost. With no larger support, the relative deficit is exactly $1/n$. The loss and gain estimates also bound total variation by the \hyperlink{nav:dgfiniteTV}{finite-class formula} in Theorem~\ref{dg:all}.

To count $\min(|A|,|Z|)=r$, take the union of the class with $r$ ones and its output complement. If $2r<s$, at least $s$ weights must be specified, so the two classes are disjoint and the count doubles exactly. If $2r=s$, the classes coincide on supports of size $s$. Larger supports contribute only $o(B_{d,r})$ to either class, so the leading term is $B_{d,r}$.
\end{proof}

At degree three, Proposition~\ref{cub:second} in Appendix~\ref{cub:sec} evaluates the first correction to $L_3$.

\section{Query costs of functions with prescribed real degree}\label{sec:jointmeasures}
Theorem~\ref{dg:all} and Proposition~\ref{dg:fixedr} give sampling rules for the whole degree class and for the class with a prescribed number of accepting weights, respectively. We use them to transfer estimates for sampled weights and labels to uniformly chosen functions.
Restricting a symmetric partial function to two weight layers with different labels gives a lower bound on each query cost. In the sparse degree range $1\ll d\ll\sqrt n$, the gap and position of the hardest pair then determine the joint limit of the two normalized costs.

\subsection{Query costs for two oppositely labelled weights}\label{nav:pairs}
A symmetric partial function of degree one is defined on exactly two weight layers with different labels (Proposition~\ref{dg:finite}). Choosing the function uniformly therefore gives equal probability to every unordered pair of weights. Write $q_n(a,b)$ for the exact quantum cost of distinguishing weights $a<b$. The two costs can be compared for a uniformly chosen pair or along a prescribed sequence of pairs.
\begin{theorem}[Query costs on two weight layers]\leavevmode\label{lj:degree}%
Fix $0<\epsilon<1/2$ and put $z=\Phi^{-1}(1-\epsilon)$, where $\Phi$ is the standard normal distribution function. Choose a degree-one function uniformly and let $n\to\infty$. Its randomized tail and its two mean query costs satisfy
\begin{equation}\label{lj:pairsummary}
 \Pr(R_\epsilon>m)\sim\frac{\pi z}{2}\sqrt{\frac{n-m}{nm}},
 \mathbb E R_\epsilon\sim\frac{\pi^2z}{4}\sqrt n,
 \mathbb E Q_E\sim\frac{\pi^2}{8}\log n.
\end{equation}
The tail estimate holds along every integer sequence $m=m(n)$ with $m\to\infty$ and $n-m\to\infty$. The expectations are over the uniform choice of function.
For a function supported on a prescribed pair $a<b$ with opposite labels, suppose $a/n\to u\in(0,1)$ and $1\ll g=b-a\ll n$. Its two query costs then satisfy
$R_\epsilon/(n-R_\epsilon)\sim(16z^2/\pi^2)Q_E^2/n$.
\end{theorem}

The denominator in the pairwise relation is positive for all sufficiently large $n$. At gap $g$, reading $n-g+1$ distinct bits leaves only $g-1$ bits unknown. The possible weights then span at most $g-1$, so at most one of the two candidates remains. Thus $n-R_\epsilon\ge g-1$, which tends to infinity in the stated interior regime.

\begin{proof}\phantomsection\label{pr:lj:degree}
Pairs not distinguishable within $m$ randomized queries determine the tail of $R_\epsilon$. Exchanging labels preserves both costs. After symmetrization and padding to $m$ distinct queries, the sample count of ones is a sufficient statistic. Given the count, all bit orderings are equiprobable under either weight. At weight $a$ this count is hypergeometric, with mean $ma/n$ and variance $m(n-m)(a/n)(1-a/n)/(n-1)$.

Set $h=\sqrt{n(n-m)/m}$. Multiplying the sample count by $n/m$ gives an estimate with standard deviation asymptotic to $h\sqrt{u(1-u)}$ when $a/n\to u\in(0,1)$. Thus $g/h$ measures the gap on the scale of this sampling error. If $m,n-m\to\infty$, $a/n\to u\in(0,1)$ and $g/h\to y\ge0$, the sample-count variances at the two weights diverge and are asymptotically equal. Measured in standard deviations, the difference between the means tends to $y/\sqrt{u(1-u)}$. The hypergeometric central limit theorem~\cite[Theorem~2.1]{lahiri} gives uniform convergence of the standardized distribution functions to $\Phi$.

Let $H_w(j)$ denote the probability of observing $j$ ones in $m$ draws at weight $w$. For positive adjacent masses, $H_w(j+1)/H_w(j)=(w-j)(m-j)/[(j+1)(n-w-m+j+1)]$ increases with $w$. The support endpoints are nondecreasing in $w$ as well, so $H_b-H_a$ changes sign at most once, from negative to positive. The total variation distance therefore equals the largest difference between the distribution functions and tends to $2\Phi(y/(2\sqrt{u(1-u)}))-1$.

The equal-prior Bayes error, half the difference between one and the total variation distance, bounds the minimax error from below. A threshold halfway between the means has error tending to $\Phi(-y/(2\sqrt{u(1-u)}))$ under each weight. It therefore attains the Bayes lower bound asymptotically even under the worst-case error criterion. \hypertarget{nav:pairtest}{Away from equality}, $m$ queries eventually fail exactly when $y<2z\sqrt{u(1-u)}$.

To count failing pairs, bound their gaps uniformly in $a$. Estimate the weight by $n/m$ times the sample count of ones. The variance bound $nh^2/[4(n-1)]$ gives midpoint error at most $nh^2/[(n-1)g^2]$, excluding gaps $g>C_\epsilon h$ from the tail count. Also $h\to\infty$ and $h/n\to0$. In coordinates $(a/n,g/h)$ the mesh is $(1/n,1/h)$. Discard strips of width $\tau$ at $u=0,1$ and around the threshold graph. The discarded cells have normalized area $O_\epsilon(\tau)+o(1)$. On the remaining compact set, failure of uniformity would give a convergent subsequence contradicting the local test. The scaled mesh contains $nh$ pairs per unit area. Divide the count below the graph by $\binom{n+1}{2}$, the number of unordered pairs. Taking $\tau\downarrow0$ gives $n\Pr(R_\epsilon>m)/h\to4z\int_0^1\sqrt{u(1-u)}\,du=\pi z/2$.

For $m=\lfloor rn\rfloor$ and $rn\ge2$, at most $O(nh)$ pairs contribute to the tail, so $\sqrt n\Pr(R_\epsilon>m)\le C_\epsilon/\sqrt r$. For $rn<2$, the bound follows since probabilities never exceed one. By the tail-sum identity,
$\mathbb E R_\epsilon/\sqrt n=\int_0^1\sqrt n\Pr(R_\epsilon>\lfloor rn\rfloor)\,dr$.
Dominated convergence and $\int_0^1\sqrt{(1-r)/r}\,dr=\pi/2$ give the randomized mean.

For a specified pair, the critical equality $g/h=2z\sqrt{u(1-u)}$ is equivalent to $m/(n-m)=4z^2u(1-u)n/g^2$. Multiply the right side by $1-\eta$ and $1+\eta$ for fixed $0<\eta<1$, and round the corresponding budgets to integers. Because $g\to\infty$ and $g=o(n)$, both the read and unread counts diverge at each budget, so rounding changes their ratio by $1+o(1)$. The smaller budget fails the local test and the larger succeeds. Letting $\eta\downarrow0$ gives $R_\epsilon/(n-R_\epsilon)\sim4z^2u(1-u)n/g^2$.

A global quantum bound controls the tails and a sharp interior estimate determines the constant. Use the two-weight Chebyshev construction of He et al.~\cite[Theorem~1 and Corollary~1]{hesun}. The degree-$2T$ polynomial $p_T(t)=[1+\cos(2T\arccos(1-2t))]/2$ maps $[0,1]$ into itself and alternates between zero and one at successive extrema. For $T=\lceil\pi n/(2g)\rceil$, put $c=n\sin(\pi/(2T))/(2g)\le1/2$. The map $l(x)=1/2+c(x-a/n)$ sends $[0,1]$ into itself and maps $a/n,b/n$ to the central extremum and the next one. Lemma~\ref{qa:real} implements $p_T\circ l$ with at most $T$ queries. Exchanging the two outputs when necessary gives $q_n(a,b)=O(n/g)$ for every pair.

For the sharper interior estimate, write $\theta(t)=\arccos(1-2t)$, $\Delta=\theta(b/n)-\theta(a/n)$. Fix $\eta>0$ and choose $T=\lceil(1+\eta)\pi/(2\Delta)\rceil$. An increasing affine map $l(x)=v+cx$ which takes $a/n,b/n$ to consecutive normalized extrema is an allowed contraction if $v\ge0$ and $v+c\le1$.

To check these inequalities, use $v=l(a/n)-ca/n$ and $c=(l(b/n)-l(a/n))/(g/n)$. Send $a/n$ to its nearest extremum in angular coordinates and $b/n$ to the next one. The inverse angular coordinate $t=(1-\cos\theta)/2$ has positive continuous derivative at $\theta(u)$. The original angular gap is $\Delta\to0$ and the new gap is $\pi/(2T)\sim\Delta/(1+\eta)$, while the starting angles differ by $O(1/T)$. Applying the mean value theorem to the inverse coordinate on both gaps gives $c\to1/(1+\eta)$. Since $l(a/n)-a/n\to0$, the two margins $v$ and $1-v-c$ tend to $\eta u/(1+\eta)$ and $\eta(1-u)/(1+\eta)$, both positive.

For the matching lower bound, $T=Q_E=o(n)$ by the upper bound. The symmetrized acceptance probability $p$ lies in $[0,1]$ throughout $[T/n,1-T/n]$ by the interval bound following~\eqref{qc:harmonic}, applied with $k=T$ to normalized weights~\cite{blek}. With $x=T/n+(1-2T/n)(1-\cos\theta)/2$, the function $q(\theta)=2p(x)-1$ has trigonometric degree at most $2T$ and satisfies $|q|\le1$ over a full period. Exchange the two labels if necessary so that the transition is from $-1$ to $1$. It takes angular distance at least $\pi/(2T)$. If the transition were shorter, we could put both transition points strictly inside one rising half-wave of a translate of $\cos(2T\theta)$. Subtract this cosine from $(1-\xi)q$. For sufficiently small $\xi>0$, the difference would have signs $+,-,+,-$ at the left endpoint, the two transition points and the right endpoint. The selected half-wave would then contain at least three zeros and each of the other $4T-1$ half-waves at least one. Their interiors are disjoint, contradicting the bound of $4T$ zeros per period.

The affine map $l(x)=(x-T/n)/(1-2T/n)$ multiplies the gap by $1+o(1)$, even if $g/n\ll T/n$. At intermediate points $x\to u$, the derivative $(\theta\circ l)'(x)=\theta'(l(x))/(1-2T/n)$ tends to $1/\sqrt{u(1-u)}$. The rescaled angular distance is therefore asymptotic to $g/(n\sqrt{u(1-u)})$. The quantum bounds match as $\eta\downarrow0$ and, with the randomized estimate, give the stated relation between the two query complexities. With $F(u)=(\pi/2)\sqrt{u(1-u)}$, the two quantum estimates are
\begin{equation}\label{lj:pairsizes}
\begin{aligned}
 q_n(a,b)&\le Cn/g &&(0\le a<b\le n,\ g=b-a),\\
 q_n(a,b)&\sim nF(u)/g &&(a/n\to u\in(0,1),\ 1\ll g\ll n).
\end{aligned}
\end{equation}

To compute the mean quantum cost, consider its tail at thresholds $1\ll j\ll n$. As for the randomized tail, summing over $(a/n,jg/n)$ gives $j\Pr(Q_E>j)\to2\int_0^1 F(u)\,du=\pi^2/8$. By \eqref{lj:pairsizes}, $jg/n$ stays bounded for pairs contributing to the tail. A positive finite limit of $jg/n$ forces $1\ll g\ll n$, so the local estimate applies off $u=0,1$. These boundaries, $jg/n=0$, and the threshold graph have zero area.

To sum these tails, use the global bound $\Pr(Q_E>j)\le C/j$. By the sequence criterion, $j\Pr(Q_E>j)$ converges uniformly on $\log n\le j\le n/\log n$. The harmonic sum there is $\log n+O(\log\log n)$. The $C/j$ bound contributes $O(\log\log n)$ on either side up to $n$, and the tail vanishes beyond $n$. Summing the tail gives the quantum mean in \eqref{lj:pairsummary}.
\end{proof}

\subsection{Typical query costs as the degree grows}
Among pairs of oppositely labelled weights, the largest $q_n(a,b)$ bounds the exact cost of a nonconstant function from below. Podder, Yao and Ye used exact tests between weights with different labels to eliminate candidates one at a time~\cite{pyyfine}. Choosing well-separated pairs early bounds the cost of eliminating all but the last pair. Poisson limits for minimum spacings of independent continuous observations are classical~\cite{kmt}. Here the points lie on the weight grid, and only pairs with opposite labels contribute.

\begin{lemma}[Distribution of the quantum cost on uniformly sampled weights]\leavevmode\label{sj:quantum}%
Choose $s$ distinct weights uniformly from $\{0,\ldots,n\}$. Independently give each sampled weight a fair binary label and leave every other weight unspecified. Suppose $s\to\infty$ and $s^2=o(n)$, and put $\Lambda=\pi^2/32$. Under this sampling law,
\begin{equation}\label{sj:sampledquantumlimit}
\begin{aligned}
 \Pr(Q_E\le ys^2)&\longrightarrow e^{-\Lambda/y} &&\text{for each fixed } y>0,\\
 \Pr(Q_E>j)&\sim\Lambda s^2/j &&\text{for integers }s^2\ll j\ll n,\\
 \mathbb E Q_E&\sim\Lambda s^2\log(n/s^2).
\end{aligned}
\end{equation}
\end{lemma}

In this sampling model the mean has an extra logarithmic factor relative to a fixed quantile. The tail estimate identifies the large costs that account for this factor.

\begin{proof}\phantomsection\label{pr:sj:quantum}
For a uniform pair $a<b$, summing \eqref{lj:pairsizes} over positions and gaps as above gives $j\Pr(q_n(a,b)>j)\to\pi^2/8$ for $1\ll j\ll n$.

\hypertarget{nav:elimination}{Let $M_*$ be} the largest cost of a pair of oppositely labelled weights, or zero if none exists. Restricting the function to that pair gives $M_*\le Q_E$. To bound the excess over $M_*$, we use inexpensive tests while at least three candidates remain. Let $L_p$ be the smallest span (last minus first weight) among sets of $p$ selected weights containing both labels, or $+\infty$ if there is no such set. Among $p\ge3$ remaining candidates carrying both labels, some pair with different labels has gap at least $L_p/2$. Use the extremes if their labels differ. Otherwise pair an oppositely labelled candidate with the farther extreme. Test the chosen pair on fresh registers. Whatever the true weight, the outcome eliminates the candidate guaranteed to give the opposite outcome. Stop when all labels agree, or test the final pair at cost at most $M_*$. To pad all branches to the worst-case query count, retain test outcomes in registers as controls and use the dummy address. By the uniform pair bound in \eqref{lj:pairsizes}, each test with $p\ge3$ candidates costs at most $2Cn/L_p$. Thus $0\le\Delta:=Q_E-M_*\le 2Cn\sum_{p=3}^s1/L_p$.

For $r\ge2$, a span $L_{r+1}\le D$ requires a selected left endpoint followed within $D$ steps by $r$ others, regardless of labels. Conditional on a selected endpoint, the chance that $r$ prescribed further sites are selected is at most $(s/N)^r$. Summing over endpoints and sites gives probability at most $s\binom{\lfloor D\rfloor}r(s/N)^r$. Setting $D=n/y$ bounds the rescaled inverse-span tail by $\min(1,(a_r/y)^r)$, where $a_r=(es/r)s^{1/r}$. Integrating from zero to $a_r$ and then to infinity gives $\mathbb E(n/L_{r+1})\le a_r r/(r-1)$. This derivation also holds without $s^2=o(n)$. Separating $r=2$ from the larger indices and summing gives $\mathbb E \Delta=O(s^{3/2}+s^{4/3}\log s)=O(s^{3/2})$.

It remains to find the distribution of $M_*$. Join weights $a,b$ by an edge when $q_n(a,b)>j$. The edge count is $(\pi^2/8+o(1))\binom N2/j$ and the maximum vertex degree is $O(n/j)$. Let $X$ count edges with oppositely labelled selected endpoints. Each of the $\binom s2$ sampled pairs has opposite labels with probability $1/2$, independently of its positions. Thus $\mathbb EX=(\Lambda+o(1))s^2/j$.

Take $j\sim ys^2$ with $y>0$ fixed. For each fixed $l$, after fewer than $l$ disjoint edges have been chosen, only $O_l(n/j)$ further edges meet a used endpoint, out of $\Theta(N^2/j)$ edges in all. Thus the number of ordered disjoint $l$-tuples is asymptotic to the $l$th power of the edge count. The probability all endpoints are sampled and each edge joins opposite labels is $(s)_{2l}/[(N)_{2l}2^l]$. Such tuples contribute $(\mathbb EX)^l(1+o(1))$ to the $l$th factorial moment of $X$.

For tuples with a shared endpoint, bound the label probability by one. Growing a spanning forest on $v$ vertices in $c$ components gives $O_l(N^c(n/j)^{v-c})$ embeddings. Multiplying this count by the probability that all its vertices are sampled gives an expected contribution of $O_l(s^{2c-v})=o(1)$ since one component has at least three vertices. The factorial moments therefore give a Poisson limit of mean $\Lambda/y$, so $\Pr(M_*\le j)=\Pr(X=0)\to e^{-\Lambda/y}$. Since $\mathbb E\Delta=o(s^2)$, the same limit holds for $Q_E$.

For the rare tail $s^2\ll j\ll n$, only the first two factorial moments are needed. Disjoint and intersecting pairs of selected edges have expected counts $O(s^4/j^2)$ and $O(s^3/j^2)$, respectively. Both are $o(\mathbb EX)$, so the first two inclusion--exclusion bounds give $\Pr(M_*>j)\sim\Lambda s^2/j$. Since $M_*$ never exceeds $Q_E$, only the exceedance upper bound needs the mean estimate for $\Delta$. An exceedance $Q_E>j$ requires $M_*>(1-\eta)j$ or $\Delta>\eta j$. Markov's inequality makes the latter probability $o(s^2/j)$ for fixed $\eta>0$. Let $\eta\downarrow0$.

For the mean we must also control the ends of the summation range. For $j\ge s^2$, the global gap bound and a union bound over the $\binom s2$ pairs give $\Pr(M_*>j/2)=O(s^2/j)$. Also $\Pr(\Delta>j/2)=O(s^{3/2}/j)$ by Markov's inequality. Thus $\Pr(Q_E>j)\le C's^2/j$ uniformly, even when $j$ is comparable to $n$. Put $H=n/s^2$ and sum the tail estimate between $s^2\exp(\sqrt{\log H})$ and $n\exp(-\sqrt{\log H})$. Uniformity follows from convergence along every such sequence. The discarded ranges contribute $O(s^2(1+\sqrt{\log H}))$. This gives the mean in \eqref{sj:sampledquantumlimit}.
\end{proof}

For $s=d+1$, independent fair labels are all equal with probability $2^{1-s}$. Conditioning on both labels occurring gives the sampling rule in Theorem~\ref{dg:all}. As $d\to\infty$, event probabilities in the independent-label model and the uniform degree-$d$ class thus differ by $o(1)$. An $o(1)$ error preserves probability bounds separated from a fixed quantile level $p$ by a positive margin. A tail of order $s^2/j\to0$ could be smaller than the error itself. The relative tail and mean estimates in~\eqref{sj:sampledquantumlimit} concern the independent-label model.

\begin{corollary}[Query quantiles in an exact degree class]\leavevmode\label{sj:budgets}%
Choose a function uniformly from the class of free real degree exactly $d$, with $d\to\infty$ and $d\le n$. Fix $0<p<1$ and $0<\epsilon<1/2$. The $p$-quantile of an integer-valued cost $X$ is the smallest $j$ for which $\Pr(X\le j)\ge p$. For $Q_E$ it has order $\Theta_p(\min(n,d^2))$, and for $R_\epsilon$ it has order $\Theta_{p,\epsilon}(\min(n,d^4))$.
If $d\gg n^{1/4}$, then $R_\epsilon/n\to1$ in probability. If $d\gg\sqrt n$, then $\Pr(Q_E\ge cn)\to1$ for some absolute $c>0$. If $d=o(n^{2/3})$, then $\Pr(Q_E>\lceil n/2\rceil+\eta n)\to0$ for every fixed $\eta>0$.
\end{corollary}

Randomized quantiles reach linear order when the degree is comparable to the fourth root of the input length. Exact quantum quantiles do so at the square-root scale.

\begin{proof}\phantomsection\label{pr:sj:budgets}
By Theorem~\ref{dg:all}, choosing $s=d+1$ weights uniformly and assigning independent fair labels changes event probabilities by $o(1)$. For the upper bounds, let $G$ be the smallest gap between opposite labels, or $\infty$ if all labels agree (probability $2^{1-s}$). For $D\ge1$, a union bound gives $\Pr(G\le D)=O(Ds^2/n)$. The \hyperlink{nav:elimination}{elimination bound} in the proof of Lemma~\ref{sj:quantum}, together with the global bound in~\eqref{lj:pairsizes}, gives $\Pr(Q_E>j)\le C s^2/j$ for every $s$ and $j\ge s^2$.

A classical algorithm samples coordinates with replacement and predicts the label of the weight nearest its estimate. An estimation error smaller than $G/2$ guarantees the correct label. By Chebyshev's inequality, $O_\epsilon(n^2/G^2+1)$ samples suffice. When $s^4<n$, take $D$ to be a sufficiently small positive constant multiple of $n/s^2$. The gap bound gives $\Pr(G>D)\ge (1+p)/2$ for large $n$. On this event the randomized cost is $O_{p,\epsilon}(s^4)$, leaving a fixed margin above $p$ to absorb the total variation error. Reading all $n$ bits caps both costs in every regime.

For the lower bounds, two close, oppositely labelled weights in the central third suffice. For $1\le D=o(n)$ let $Z_D$ count pairs of oppositely labelled sampled weights at distance at most $D$ inside $[n/3,2n/3]$. Its mean is $\asymp Ds^2/n$. Indicators of disjoint edges have nonpositive covariance by sampling without replacement. Edge pairs sharing a vertex contribute $O(D^2s^3/n^2)$ to the variance. Hence $\Pr(Z_D=0)\le1/\mathbb E Z_D+O(1/s)$. Choose $D=\lceil C_p n/s^2\rceil$ with $C_p$ large enough that this probability is below $p/2$. With probability greater than $1-p/2$, at least one such pair forces the cost lower bound needed for the $p$-quantile. Each such pair requires at least $cn/D$ exact queries. If its cost $T$ is at least $n/10$, this follows from $D\ge1$. Otherwise the central third lies strictly inside $[T/n,1-T/n]$ in normalized coordinates. On that third the rescaled angular map in Theorem~\ref{lj:degree} has bounded derivative, so the angular gap is at most $C D/n$. The half-wave bound requires this gap to be at least $\pi/(2T)$ even for a one-site separation.

For the randomized cost let $m=\lfloor c'\min(n,s^4)\rfloor$. Both $m,n-m$ diverge, and $D/\sqrt{n(n-m)/m}\le C_p\sqrt{c'/(1-c')}$ for all sufficiently large $n$, with $C_p$ depending only on $p$. By the \hyperlink{nav:pairtest}{two-weight test} in the proof of Theorem~\ref{lj:degree}, for small ratios the limiting minimax error for the two hypergeometric laws is arbitrarily close to $1/2$, uniformly on the central third by compactness. For small $c'$, the pair needs more than $m$ randomized queries. This proves the lower bounds on both quantiles.

If $s\gg\sqrt n$, taking $D=1$ gives two adjacent, oppositely labelled weights in the central third with probability tending to one. The pairwise lower bound then gives linear quantum cost. If $s\gg n^{1/4}$, choose an integer $D=o(\sqrt n)$ with $Ds^2/n\to\infty$. By the preceding second-moment bound, the central third contains a pair of opposite labels within this distance with probability tending to one. For fixed $0<b<1$, the two-layer error at $\lfloor bn\rfloor$ reads tends uniformly to $1/2$ for these pairs, proving $R_\epsilon/n\to1$.

To obtain the sharper quantum upper bound for $s=o(n^{2/3})$, it suffices to bound the final pair uniformly by $T=\lceil n/2\rceil$ queries. One phase query on $(|i\rangle+|j\rangle)/\sqrt2$ gives the parity of the two addressed bits by a measurement in the sum and difference basis. Pairing the bits and using the dummy address when one bit remains gives total parity in $T$ queries, which distinguishes adjacent weights.

For gap $g\ge2$ use adjacent Chebyshev extrema as in the proof of Theorem~\ref{lj:degree}. Consecutive normalized degree-$2T$ nodes are at most $\pi/(4T)$ apart. Choose a node within $O(1/n)$ of $1/10+(4/5)a/n$ and its right neighbour. The affine map sending $a/n,b/n$ to these nodes has slope at most $\pi/(2g)\le\pi/4<4/5$. Writing $l$ for this map and $c$ for its slope gives $l(0)=1/10+(4/5-c)a/n+O(1/n)$ and $l(1)=l(0)+c\le9/10+O(1/n)$. Both endpoint constraints thus hold uniformly over pairs. Composing the corresponding scaled Chebyshev probability with this map gives degree $2T$, range $[0,1]$ and opposite labels at the two specified weights. Lemma~\ref{qa:real} implements it in $T$ queries. Thus $M_*\le\lceil n/2\rceil$. The elimination tests before the final pair have total expected cost $O(s^{3/2})=o(n)$. Markov's inequality and the total variation bound in Theorem~\ref{dg:all} give the asserted bound in probability on the degree-$d$ class.
\end{proof}

\subsection{Joint limits from the positions and gaps of oppositely labelled weights}\label{sj:jointsection}

Two pairs of oppositely labelled weights can have the same gap and different costs. For example, take $1\ll g\ll n$ and place one pair at $a/n\to1/2$ and the other at $a/n\to1/10$, with gap $g$ in each case. The local estimate in \eqref{lj:pairsizes} gives a ratio of exact costs tending to $(1/2)/(3/10)=5/3$.

For weights $t,t+g$, put $u=t/n$. The score $\sqrt{u(1-u)}/g$ increases as the gap narrows at a fixed interior location. At a fixed gap it increases as the smaller weight approaches the centre. At an interior weight, the standard deviation of the sample proportion after $1\le m<n$ reads is asymptotic to $\sqrt{u(1-u)}/\sqrt{nm/(n-m)}$. Comparing this sampling error with the difference $g/n$ between the two proportions gives the same dependence on gap and location.

Sampling $r$ accepting and $k$ rejecting weights gives $rk$ pairs with different labels. In the sparse regime below, the scale of the smallest gaps is $n/(rk)$. In the limits below, $\Rightarrow$ denotes convergence in distribution.

\begin{theorem}[Joint query complexities with a prescribed number of accepting weights]\leavevmode\label{sj:colours}%
Choose a function uniformly from the degree-$d$ class with exactly $r$ accepting weights. Put $s=d+1$, $k=s-r$ and $b=rk$. Under the sampling rule in Proposition~\ref{dg:fixedr}, $s$ is the number of selected weights, $k$ counts those labelled zero and $b$ counts pairs with different labels. A function in the degree class can have more than $s$ specified weights.
Suppose $r\to\infty$, $r\le k$ and $b=o(n)$. Fix $0<\epsilon<1/2$ and set $z=\Phi^{-1}(1-\epsilon)$. The randomized cost satisfies
\begin{equation}\label{sj:fixedcolourlimit}
 \frac1b\sqrt{\frac{nR_\epsilon}{n-R_\epsilon}}\ \Rightarrow\ 2zV,
 \qquad \Pr(V\le v)=\exp[-\pi/(4v)]\quad(v>0).
\end{equation}
Write $Z_n$ for the normalized cost on the left, taking $Z_n=+\infty$ when $R_\epsilon=n$. Among the pairs of weights $t<t+g$ with different labels in the sampled function, choose one maximizing $\sqrt{(t/n)(1-t/n)}/g$, with arbitrary tie-breaking. Let $U_n=t/n$ for the chosen pair. Then $(U_n,Z_n)\Rightarrow(U,2zV)$, where $U$ is independent of $V$ and has density $8\sqrt{u(1-u)}/\pi$ on $(0,1)$.
Under the additional condition $\log s=o(r)$, the two costs satisfy $(Q_E/b,Z_n)\Rightarrow((\pi/2)V,2zV)$.
\end{theorem}

A maximum-score pair is a candidate for the hardest pair. Under the additional growth condition, its rescaled score governs both limiting costs. For the whole promise, we also control nearby alternatives in the randomized algorithm and count quantum queries before the final pair test.

\begin{proof}\phantomsection\label{pr:sj:colours}
By Proposition~\ref{dg:fixedr}, the uniform law in the theorem differs by $o(1)$ in total variation from sampling $r$ accepting and $k$ rejecting weights without replacement. Set $H=n/b$ and record each pair of oppositely labelled weights $t,t+g$ as $(u,y)=(t/n,g/H)$. A bounded rectangle of area $A$ in $[0,1]\times(0,\infty)$ contains $(A+o(1))nH$ possible pairs of grid positions. Each is oppositely labelled with probability $2rk/[(n+1)n]$, so the expected count tends to $2A$.

For disjoint rectangles with bounded $y$, the mixed factorial moments of their point counts sum over ordered tuples of distinct edges, each edge joining weights with different labels. Suppose a tuple uses $i$ accepting and $j$ rejecting vertices in $c$ connected components. Its prescribed labels occur with probability $(r)_i(k)_j/(n+1)_{i+j}$. Choose the first vertex of each component freely. Bounded path lengths place each vertex within $O(H)$ of its component's first vertex, giving $O(n^cH^{i+j-c})$ placements. The placement bound times the label probability gives $O(r^{c-j}k^{c-i})$. Each component contains both labels, so $i,j\ge c$. A shared endpoint makes at least one inequality strict, and the contribution vanishes as $r,k\to\infty$. For disjoint edges, the joint inclusion probability is asymptotic to the product of the one-edge probabilities for any fixed tuple length. The Riemann sums then give the product of the limiting rectangle means. These moments identify a Poisson process of intensity $2\,du\,dy$ on bounded gap ranges.

Let $V_n$ be the largest rescaled score $v=\sqrt{u(1-u)}/y$. Since $v=H\sqrt{u(1-u)}/g$, rescaling preserves the maximizing pair. In $(u,v)$ coordinates the limiting intensity is $2\sqrt{u(1-u)}\,du\,dv/v^2$. Its mass above a score $v>0$ is $\pi/(4v)$ since the integral of $\sqrt{u(1-u)}$ over $[0,1]$ is $\pi/8$. Thus the limiting process has no point above $v$ with probability $\exp[-\pi/(4v)]$.

To pass from bounded rectangles to the maximizing pair, first keep scores above $c>0$ and gaps $y\ge\alpha>0$. Then $\alpha\le y\le1/(2c)$, and the largest retained score and its location converge jointly, assigning $(0,0)$ to an empty configuration. Cutoff boundaries have zero limiting intensity and the limiting scores are almost surely distinct. The expected number of observed pairs with $y<\alpha$ is $O(\alpha+1/H)$, whereas the limiting mean is $2\alpha$. We may therefore let $\alpha\downarrow0$. The probability of an empty retained set now tends to $\exp[-\pi/(4c)]$, which vanishes as $c\downarrow0$. The unrestricted maximum and its location therefore converge jointly, and an arbitrary choice among tied maximizers gives the same limit. The intensity factors into a function of $v$ and the location density $8\sqrt{u(1-u)}/\pi$. Conditional on the scores, each location has that density. The location of the largest score is independent of $V$.

For the randomized cost, fix $w>0$ and let $m=\lfloor wnb^2/(n+wb^2)\rfloor$. Then $R_\epsilon\le m$ is exactly the event $Z_n\le\sqrt w$. Before rounding, the read and unread budgets are $wnb^2/(n+wb^2)$ and $n^2/(n+wb^2)$. Both diverge since $b\to\infty$ and $b=o(n)$, so $nm/[(n-m)b^2]=w(1+o(1))$. Estimate the weight by $n$ times the sample proportion. The estimator has variance at most $(1+o(1))H^2/(4w)$. On $\alpha\le y\le L$, $\delta\le u\le1-\delta$, the \hyperlink{nav:pairtest}{two-weight test} in the proof of Theorem~\ref{lj:degree} gives the minimax error $\Phi(-y\sqrt w/[2\sqrt{u(1-u)}])+o(1)$ uniformly. Each midpoint error has the same estimate. In score coordinates the leading error is $\Phi(-\sqrt w/(2v))$, which reaches $\epsilon$ at $v=\sqrt w/(2z)$.

A two-weight test gives a lower bound by restriction. For an upper bound on the entire promise, we bound the probability that a selected weight has at least two nearby weights with the opposite label. The expected number of oppositely labelled pairs satisfying $y<\alpha$ is $O(\alpha+1/H)$, and for $y\le L$ with $u$ within $\delta$ of either endpoint it is $O(L\delta)+o(1)$. Three selected weights spanning at most $2LH$ carry both labels only if two are accepting and one rejecting or conversely. The two possibilities give a total expected count $O_L(bsH^2/n^2)=O_L(1/r+1/k)=o(1)$.

Outside these exceptions, each selected weight has at most one weight with the other label within $LH$. Predict the label of the weight nearest the estimate. To choose a nearby weight with the other label the estimate must reach or cross their midpoint. To choose any weight farther than $LH$ from the true weight the estimate must err by more than $LH/2$, an event of probability $O_w(L^{-2})$ by the variance bound. If $V_n\le\sqrt w/(2z)-\eta$, the midpoint errors lie below $\epsilon$ by a fixed margin that absorbs the error from distant weights for large $L$. If $V_n\ge\sqrt w/(2z)+\eta$, the optimal error for the maximizing pair alone exceeds $\epsilon$. Here $L$ can be chosen to include any pair above this positive threshold since $v\le1/(2y)$.

Choose $L$ large and then $\alpha,\delta$ small. Continuity of the law of $V$ allows $\eta\downarrow0$. Thus the events $R_\epsilon\le m$ and $2zV_n\le\sqrt w$ disagree with probability tending to zero. Comparison on arbitrarily fine finite grids in a bounded interval of arbitrarily high probability gives $Z_n-2zV_n\to0$ in probability. Together with the joint limit of the maximizing pair, this proves the claim about $(U_n,Z_n)$. In particular the probability of an infinite value vanishes.

For quantum queries, first consider the largest cost $M_*$ among pairs with different labels. For $g>LH$, the global estimate in \eqref{lj:pairsizes} and the bound on the score give $q_n/b$ and $(\pi/2)\sqrt{u(1-u)}/y$ the common upper bound $\max(C,\pi/4)/L$. For $g\le LH$, the probability that some oppositely labelled pair has $y<\alpha$ or $u\notin[\delta,1-\delta]$ is $O(\alpha+L\delta)+o(1)$. On the remaining compact range the local estimate in \eqref{lj:pairsizes} makes the difference of the two quantities $o(1)$ uniformly. Taking maxima, then choosing $L$ large and $\alpha,\delta$ small, yields $M_*/b-(\pi/2)V_n\to0$ in probability.

It remains to pass from $M_*$ to the cost of all specified weights. For $p\ge3$ let $L_p$ be the \hyperlink{nav:elimination}{minimum span of $p$ selected weights containing both labels}, with $1/(+\infty)=0$. The algorithm in Lemma~\ref{sj:quantum} finds two candidates with different labels at distance at least $L_p/2$ and tests them to eliminate one candidate. The final pair costs at most $M_*$. Restriction and this algorithm therefore give $M_*\le Q_E\le M_*+Cn\sum_{p=3}^s L_p^{-1}$.

If $L_{j+1}\le D$, some accepting weight has at least $j$ other selected weights within distance $D$. Index the $r$ accepting weights in a uniformly random order. Conditional on the position of any one of them, the other $s-1$ selected weights form a uniform subset of the remaining sites. A union bound over the $r$ indices gives
$\Pr(L_{j+1}\le D)\le\min\{1,r(CsD/(jn))^j\}$.
Integrating the resulting tail bound yields $\mathbb E(n/L_{j+1})\le C'(s/j)r^{1/j}$ for $j\ge2$. Sum separately at $j=2$, from $3$ to $\max(3,\lceil\log r\rceil)$ and above this threshold. Summing over these ranges bounds the expected extra cost by a constant times
$s[\sqrt r+r^{1/3}\log\log(r+3)+\log s]$. Since $b=rk\ge rs/2$, the first two terms are $o(b)$ as $r\to\infty$. The last is $o(b)$ by $\log s=o(r)$. Markov's inequality now gives $(Q_E-M_*)/b\to0$ in probability. The normalized quantum and randomized costs thus equal $(\pi/2)V_n$ and $2zV_n$, respectively, up to errors tending to zero in probability. This gives their joint limit.
\end{proof}

The preceding theorem conditions on the number of accepting weights. Averaging over that count gives the joint law on the sparse degree class.

\begin{corollary}[Joint query complexities on the entire sparse degree class]\leavevmode\label{sj:jointlaw}%
Choose a function uniformly from the whole degree-$d$ class, where $d\to\infty$ and $d^2=o(n)$. Fix $0<\epsilon<1/2$ and put $z=\Phi^{-1}(1-\epsilon)$. Taking the randomized coordinate to be $+\infty$ if $R_\epsilon=n$, we have
\begin{equation}\label{sj:sparsejointlimit}
 \left(\frac{Q_E}{d^2},\frac{nR_\epsilon}{d^4(n-R_\epsilon)}\right)
 \ \Rightarrow\ \left(Y,\frac{16z^2}{\pi^2}Y^2\right),
\end{equation}
where $\Pr(Y\le y)=e^{-\pi^2/(32y)}$ for $y>0$. In particular, $nR_\epsilon/[(n-R_\epsilon)Q_E^2]\to16z^2/\pi^2$ in probability.
If also $d^4/n\to\gamma\in(0,\infty)$, then for each fixed $0<a<1$ the number of degree-$d$ functions with $R_\epsilon\le an$ is asymptotic to $L_d\exp[-(\pi z/8)\sqrt{\gamma(1-a)/a}]$, where $L_d$ is given in Theorem~\ref{dg:all}.
\end{corollary}

When the ratio of $d^4$ to $n$ tends to a finite positive limit, the fraction of the degree class computable within $an$ randomized queries converges. Its limit depends on the fixed budget fraction $a$, the limiting ratio $\gamma$ and the error tolerance.

\begin{proof}\phantomsection\label{pr:sj:jointlaw}
By Theorem~\ref{dg:all} the uniform degree-$d$ law is within $o(1)$ in total variation of choosing $s=d+1$ weights uniformly and giving them a uniform nonconstant labelling. Independent fair labels add at most $2^{1-s}$, the probability of a constant labelling, to the total variation error. Let $r$ be the smaller label count. Exchanging the outputs preserves both query costs, so we may name the less frequent label one, keeping the original names in a tie. Conditional on $r$, the $r$ ones are then uniform on the $s$ selected positions, exactly as in the sampling argument for Theorem~\ref{sj:colours}.

With probability tending to one, $s/2-s^{2/3}\le r\le s/2$ by the binomial variance bound. Every sequence of counts in this window satisfies $r/s\to1/2$, $b=r(s-r)=o(n)$ and $\log s=o(r)$. The conditional limits thus hold uniformly on the window for each bounded continuous test function. Indeed, a sequence of counts violating uniformity would itself satisfy the same hypotheses. Averaging over $r$ and using $b/d^2\to1/4$ in probability now gives the unconditional limit.

The quantum coordinate has limit $Y=(\pi/8)V$. On the typical window the randomized coordinate is the square of $(b/d^2)Z_n$, with $Z_n$ as in Theorem~\ref{sj:colours} conditional on $r$, so its limit is $(zV/2)^2=(16z^2/\pi^2)Y^2$. This gives~\eqref{sj:sparsejointlimit}. Since $Y>0$ almost surely, dividing by the squared quantum coordinate gives convergence to a constant in distribution and hence in probability.

Finally $R_\epsilon\le an$ is equivalent to the randomized coordinate being at most $an/[d^4(1-a)]$. When $d^4/n\to\gamma$, the limiting event for $Y$ is $Y\le\frac{\pi}{4z}\sqrt{a/[\gamma(1-a)]}$. The cutoff is a continuity point of the limiting distribution function. Substituting this cutoff into the distribution function of $Y$ and multiplying by the degree-class size $(1+o(1))L_d$ gives the count.
\end{proof}

For $1\ll d\ll\sqrt n$, setting the limiting distribution function in Corollary~\ref{sj:jointlaw} equal to $1/2$ gives a median exact quantum cost asymptotic to $\pi^2d^2/(32\log 2)$. Prescribing the label counts can change even the order of a quantile. If $1\ll d\ll n^{1/4}$, fixed interior quantiles of $Q_E$ and $R_\epsilon$ have respective orders $(d^{4/3},d^{8/3})$ for $r=\lfloor d^{1/3}\rfloor$ and $(d^2,d^4)$ for $r=\lfloor(d+1)/2\rfloor$. In both cases $b^2=o(n)$, so the randomized normalization in~\eqref{sj:fixedcolourlimit} is equivalent to $\sqrt{R_\epsilon}/b$ in probability.

\section{Open counting questions}
Theorem~\ref{ov:qcount} determines the exponent of $n$ at every fixed positive exact quantum cost $k$. For fixed $k\ge2$, does the number of functions with $Q_E=k$, divided by $n^k$, converge as $n\to\infty$? If so, what is the limit?

For fixed $0<\epsilon<1/2$ and $1\ll T\ll n$, consider symmetric partial functions with $Q_E=T$. Beyond the range $1\le r=o(\min\{T^2,\sqrt n\})$ in Theorem~\ref{rd:doublebudget}, which values of $R_\epsilon$ occur, and how many of these functions attain each value?

At fixed randomized cost, Theorem~\ref{ra:fixed} gives the exponential growth rate from the minimum gap between boundary fractions of ones. How does rounding the two boundary weights outwards affect the count beyond this growth rate? For counts by free real degree, Proposition~\ref{cub:second} gives the first correction to $L_3$. What is the first correction at higher fixed degrees?

\section*{Statements and Declarations}

\noindent Data and code availability.
This theoretical study uses no empirical dataset.

\noindent Competing interests.
The authors declare no competing interests.

\noindent Funding.
This work is supported in part by the National Natural Science Foundation of China (Nos. 62272208 and 62572221).

\noindent Author contributions.
G.X. was responsible for conceptualization, formal analysis, supervision, and funding acquisition.
G.X., M.L., and M.W. contributed to methodology, investigation, preparation of the original draft, and visualization.
M.W., and H.Z. performed validation.
G.X., M.L., M.W., and H.Z. reviewed and edited the manuscript.
All authors have read and approved the final manuscript.

\noindent\emph{Use of generative AI.}\par\nopagebreak[4]
OpenAI ChatGPT was used for mathematical exploration, literature searches, manuscript preparation and verifier development. The authors take responsibility for the results and final manuscript.

\ifjournalstyle\else\newenvironment{appendices}{\appendix\setcounter{equation}{0}\renewcommand{\theequation}{\Alph{section}\arabic{equation}}}{}\fi
\begin{appendices}
\renewcommand*{\theHequation}{\theequation}

\section{Proofs of the counts for two randomized queries}\label{app:twoproofs}
With two randomized queries the symmetrized acceptance probability has degree at most two in the weight, so labels change at most twice. For two changes, orient the middle block as accepting. Concavity then confines its weights to the central window $W$ of Theorem~\ref{rtwo:turns}. For one change oriented from zero to one, the last zero and first one determine feasibility through the thresholds $b_q(a)$ of Section~\ref{rtwo:sec}. The second proof adds these disjoint counts.
\begin{proof}[Proof of Theorem~\ref{rtwo:turns}]\phantomsection\label{pr:rtwo:turns}
For part (i), orient the labels as zero, one, zero. At weight $t$ write $P_i(t)=\binom ti\binom{n-t}{2-i}/\binom n2$ for the probability of seeing $i$ ones in two reads. By \eqref{ra:sampling}, the acceptance probability is $B(t)=\sum_{i=0}^2 u_iP_i(t)$ for some $0\le u_i\le1$. To reach $2/3$ between two values at most $1/3$, $B$ must be a strictly concave quadratic. Its vertex lies between the two outer rejecting weights. Beyond them $B$ decreases towards either endpoint, giving $u_0=B(0)\le1/3$ and $u_2=B(n)\le1/3$. At any accepting weight $c$ we have
$2/3\le B(c)\le1/3+(2/3)P_1(c)$. Thus $P_1(c)=2c(n-c)/(n(n-1))\ge1/2$, equivalently $(2c-n)^2\le n$, which gives $c\in W$. Conversely, $(u_0,u_1,u_2)=(1/3,1,1/3)$ works when the only rejecting weights are $0,n$ and the accepting weights form any nonempty subset of $W$. A one-query acceptance probability is affine, so it cannot cross one half twice.

It remains to locate the rejecting weights. Suppose $n>4$ and a left zero has weight at least two. On the interval from zero to that weight, concavity and the later rise to an accepting weight $c$ give $B(2)\le1/3$ and $c>2$. Subtract $P_0(c)B(2)/P_0(2)$ from $B(c)$ to eliminate $u_0$. Both remaining coefficients are nonnegative because $P_i(t)/P_0(t)$ increases with $t$ for $i=1,2$ (cross-multiply if $P_0(c)=0$). Using $u_1\le1$ and $u_2\le1/3$ in these two terms gives $B(c)\le \tfrac13+\tfrac23(P_1(c)-4P_0(c)/(n-3))$. Writing $j=c-n/2$, the required inequality $B(c)\ge2/3$ would imply
\begin{equation}\label{rtwo:interiorzero}
 (n-3)(n-4j^2)-8(n/2-j)(n/2-j-1)=-(n-1)[4(j-1)^2+n-4]\ge0.
\end{equation}
The factored expression in~\eqref{rtwo:interiorzero} is strictly negative for $n>4$. Hence the left zeros lie in $\{0,1\}$ and, by reflection, the right zeros in $\{n-1,n\}$.

The two interior choices $1$ and $n-1$ cannot both be zeros. If they were, each of $u_0,u_2$ would be at most $(n/3-2u_1)/(n-2)$. In the resulting upper bound on $B(c)$ the coefficient of $u_1$ is $(nP_1(c)-2)/(n-2)>0$, since $c\in W$ and $n>4$. Even at $u_1=1$, reaching the acceptance threshold would require $P_1(c)\ge1/2+1/n$, whereas $P_1(c)\le n/[2(n-1)]$.

Consider functions with an interior rejecting weight. After reflection the rejecting weights are therefore $1,n$ and optionally $0$. From $B(1)\le1/3$ obtain $(n-2)u_0+2u_1\le n/3$. In the resulting upper bound on $B(c)$, the coefficient of $u_1$ is $P_1(c)-2P_0(c)/(n-2)>0$ at every accepting weight $c>1$. For $n\ge6$ every such $c$ has the same maximizer, $u_1=1$, $u_2=1/3$, $u_0=(n-6)/(3(n-2))$. At this maximizer, $B(c)\ge2/3$ exactly when the midpoint offset satisfies $j(1-j)\ge0$. The accepting weights can therefore be any nonempty subset of $[n/2,n/2+1]\cap\mathbb Z$. The two choices at $0$, two orientations under reflection and two output labellings give eight functions per accepting set. For even $n$ there are three nonempty accepting sets, giving $24$ functions. For odd $n$ there is one, giving $8$.

For $n=2$ two changes already prescribe both endpoints. For $n=3,4,5$, if either outer interval contains an interior zero, reflect if necessary to place one on the left. Then $B(1)\le1/3$ and $u_2\le1/3$. At each possible $c\in W$, $c\ge2$, eliminating $u_1$ gives $B(c)\le nP_1(c)/6+P_2(c)/3$, since the residual coefficient of $u_0$ is nonpositive. The respective maxima on these three grids are $4/9,1/2,3/5$, all below $2/3$.

For part (ii), take $0,n$ as the only rejecting weights and a nonempty $S\subseteq W$ as the accepting weights. Write $y=|x|/n$ and $F(y)=\prod_{t\in S}(1-ny/t)$. For the free degree, complement the labels. In the normalized variable, the interpolant must equal one at both endpoints and vanish at the $s$ normalized weights corresponding to $S$. At degree $s$, these roots and the value at zero determine $F$ uniquely. Its value at one is $(-1)^sD_S$. If this differs from one, interpolation on the $s+2$ specified weights gives degree exactly $s+1$.

For the quantum lower bound, fix the weight-zero final state. On each layer, average the real part of its inner product with the input-dependent final state, as in Lemma~\ref{ax:signal}. The resulting polynomial has degree at most the query count. It has the same $s$ roots, equals one at zero and has modulus at most one at every grid point. A cost of $s$ therefore forces this polynomial to be $F$ and gives $D_S\le1$. Starting at weight $n$ gives the reflected inequality $D_S^{-1}\le1$. Hence a cost of $s$ requires $D_S=1$.
For the matching algorithm, reflect the input if needed so that $D_S\ge1$. The affine correction in $H(y)=F(y)(1-y+y/D_S)$ preserves the zeros and makes $|H(0)|=|H(1)|=1$. To apply Lemma~\ref{qa:real} to $1-H^2$, we must bound $|H|$ on the whole interval.

For $n\ne4$ all normalized roots lie in $[1/3,2/3]$, by $\sqrt n\le n/3$ for $n\ge9$ and direct inspection for smaller $n$. Left of the roots, $|F|\le1$. Between the extreme roots, each factor $|1-ny/t|\le1$ because the largest root is at most twice the smallest. The affine correction is positive and at most one. To the right of the roots the logarithmic derivative of $|F|$ is at least $s$, and that of the correction factor is at least $-(D_S-1)$. Thus $D_S-1\le s$ will make $|H|$ nondecreasing up to its endpoint value one.

Still excluding the four-bit grid, a singleton has normalized root at least $1/3$, giving $D_S\le2$. For $n\le8$, $n\ne4$, at most one root lies below the middle weight. Its factor in $D_S$ is at most two and every other factor is at most one, so $D_S-1\le1$. For $n\ge9$, include every factor from a below-midpoint weight in the central window and discard all other factors to obtain
\begin{equation}\label{rtwo:logproduct}
 \log D_S\le\frac2{n-1}
 \sum_{\substack{1\le x\le\sqrt n\\ x\equiv n\ (2)}}x
 \le\frac{(\sqrt n+1)^2}{2(n-1)}\le1.
\end{equation}
Here $x=n-2t$ measures displacement towards the lower weights, and $\log[(n+x)/(n-x)]=2\int_0^{x/n}(1-v^2)^{-1}\,dv\le2x/(n-1)$ since $x^2\le n$. The grid sum is bounded by $(\sqrt n+1)^2/4$ for either parity. Thus \eqref{rtwo:logproduct} gives $D_S-1\le e-1<2$ for $s\ge2$. This proves the derivative bound on the last interval and hence $|H|\le1$ everywhere. Lemma~\ref{qa:real} implements the acceptance probability $1-H(y)^2$ in at most $\deg H$ queries, with $H=F$ if $D_S=1$.
The remaining grid is $n=4$, where $W=\{1,2,3\}$. After one phase query on $(|i\rangle+|j\rangle)/\sqrt2$, measure in the sum and difference basis to recover $x_i\oplus x_j$. For nonempty $S\subseteq\{1,3\}$, two pair-parity queries compute parity. For $S=\{2\}$ one uniform phase query suffices. For all remaining $S$, three pair-parity queries test whether all bits are equal. The root and product lower bounds give equality in every case.

By \eqref{rtwo:endpointcost}, the two costs can differ only when $D_S=1$ and $s$ is odd. To locate these sets, use the signed displacement $x_t=2t-n$, now positive above the midpoint. Taking logarithms in $D_S=1$ gives $\sum_{t\in S}\operatorname{arctanh}(x_t/n)=0$. Isolating the linear term in the inverse hyperbolic tangent expansion gives
\begin{equation}\label{rtwo:sumzero}
 \left|\sum_{t\in S}x_t\right|\le
 \frac{\sum_{t\in S}|x_t|^3}{3n^2(1-1/n)}
 \le\frac{1+1/\sqrt n}{3(1-1/n)}<1\quad(n\ge3).
\end{equation}
The integer sum in \eqref{rtwo:sumzero} is therefore zero (also at $n=2$, where $W=\{1\}$). In odd dimension each $x_t$ is odd, so every $S$ with $D_S=1$ has even size. In even dimension, any union of reflected pairs plus the middle weight gives a set of odd size with $D_S=1$. There are $(|W|-1)/2$ reflected pairs to choose from, which proves the claimed multiplicity in even dimensions.

For part (iii), write $A_i$ for the number of nonempty sets $S$ of size $i$ with $D_S=1$, as in Section~\ref{rtwo:sec}. Each weight in $W$ has its reflection there, so $A_{|W|}=1$. For $1\le i<|W|$, reveal $i-1$ elements of a uniform $i$-set. At most one of the $|W|-i+1$ remaining weights makes the product $D_S$ equal one. Since $D_W=1$, the complement $W\setminus S$ also has product one whenever $D_S=1$. Applying the same bound to the complement gives $A_i\le\binom{|W|}{i}/(1+\max(i,|W|-i))$. Also $A_1\le1$ and $A_2=\lfloor |W|/2\rfloor$.

For a fixed size $i\ge3$ we need a sharper bound. Eliminate $x_i$ using $\sum x_t=0$ in $\prod(n+x_t)=\prod(n-x_t)$. Evaluating at the sum-zero tuple $(1,1,-2,0,\dots,0)$ shows that the polynomial is not identically zero. Repeated entries are allowed in this test, and the grid bound covers the distinct tuples being counted. A nonzero bounded-degree polynomial in $i-1$ variables has $O_i(|W|^{i-2})$ grid zeros, by induction on the variables using a nonzero leading coefficient. Hence $A_i=O_i(|W|^{i-2})$. By \eqref{rtwo:endpointcost}, a set of size $k-1$ with $D_S\ne1$ contributes to $Q_E=d=k$, as does a set of size $k$ with $D_S=1$ when $k$ is even. Doubling for output complementation gives
$2[\binom{|W|}{k-1}-A_{k-1}+\mathbf1_{\{2\mid k\}}A_k]$ for $2\le k\le |W|$.
For fixed $k\ge3$, $A_k=O_k(|W|^{k-2})$ and $A_2=\lfloor |W|/2\rfloor$, so the corrections are $o(\binom{|W|}{k-1})$. For growing $k<|W|$, the complementary bound gives
\begin{equation}\label{rtwo:growingcorrections}
 \frac{A_{k-1}}{\binom{|W|}{k-1}}\le\frac{2}{|W|},
 \frac{A_k}{\binom{|W|}{k-1}}\le\frac{|W|-k+1}{k(1+\max(k,|W|-k))}\le\frac1k.
\end{equation}
At $k=|W|$ the terms $A_{|W|-1}=A_1\le1$ and $A_{|W|}=1$ are negligible relative to $\binom{|W|}{|W|-1}=|W|$. Together with the fixed-size and endpoint estimates,~\eqref{rtwo:growingcorrections} covers every sequence $3\le k\le |W|$ as $|W|\to\infty$, proving uniformity. At $k=2$, the expression is $2(|W|-A_1+\lfloor |W|/2\rfloor)=3|W|+O(1)$. The $O(1)$ exceptional functions do not affect these estimates since $\binom{|W|}{k-1}\ge |W|$ for $2\le k\le |W|$.

Choose uniformly among the functions constructed from $S$ in part (i) and their output complements. Then $S$ is uniform over the nonempty subsets of $W$, and $\Pr(D_S=1)=O(1/|W|)$ by the bound on $A_i/\binom{|W|}{i}$ for $i<|W|$ and the single remaining set at $i=|W|$. Hence $Q_E=d$ with probability tending to one. The size $s$ has distribution $\mathrm{Bin}(|W|,1/2)$ conditioned on $s>0$. Conditioning discards probability $2^{-|W|}$, and replacing $s$ by either cost changes the standardized value by at most $2/\sqrt{|W|}$. The binomial central limit theorem gives both normal limits. Passing to the full two-change class adds an exceptional fraction $e_n/[2(2^{|W|}-1)+e_n]=O(2^{-|W|})$.
\end{proof}

\begin{proof}[Proof of Corollary~\ref{rtwo:all}]\phantomsection\label{pr:rtwo:all}
Consider a single change, orienting the labels so that the $r$ zero weights $a_i$ precede the $s$ one weights $b_j$. To count by degree we first show that $d=r+s-1$. By Section~\ref{ov:model} it suffices to work with the unique interpolant of degree at most $r+s-1$. Factor out its $r$ prescribed roots, writing it as $\prod_i(t-a_i)h(t)$. Then $h$ has degree at most $s-1$ and agrees at the $b_j$ with $g(t)=\prod_i(t-a_i)^{-1}$.

The coefficient of degree $s-1$ in $h$ cannot vanish. To the right of all $a_i$, the Leibniz rule gives $(-1)^k g^{(k)}>0$, since every term has that sign. If $s>1$, applying Rolle's theorem $s-1$ times to the $s$ zeros of $g-h$ yields $h^{(s-1)}=g^{(s-1)}\ne0$ somewhere between $b_1$ and $b_s$. If $s=1$, $h=g(b_1)>0$ is constant. Both cases give the claimed degree.

Fix the last zero $a$ and the first one $b$. The monotone threshold tests of Section~\ref{rtwo:sec} meet the error bound at all exterior specified weights whenever they meet it at $a,b$. Hence the randomized query complexity equals two precisely for $b_2(a)\le b<b_1(a)$. The $n+a-b$ exterior weights are optional. With $j-1$ of them specified there are $j+1$ specified weights in total, hence degree $j$. Thus these boundary weights contribute $2^{n+a-b}$ functions overall and $\binom{n+a-b}{j-1}$ of degree $j$. As $b$ runs from $b_2(a)$ to $b_1(a)-1$, the geometric sum and the identity $\sum_{h=L}^{U}\binom h{j-1}=\binom{U+1}j-\binom Lj$ give the first terms in \eqref{rtwo:totalcount} and \eqref{rtwo:degreecount}. The factor two counts both orientations. Equal thresholds give an empty range, and $b_1(a)=n+1$ allows $b=n$.

For two changes with rejecting endpoints, start by assigning degree $|S|+1$ to every nonempty accepting set. By \eqref{rtwo:endpointcost} a set with $D_S=1$ and even size moves down one degree. Among sets with product one, this removes those of size $j-1$ if $j-1$ is even and adds those of size $j$ if $j$ is even. Output complementation doubles the bracket's count of functions with rejecting endpoints. At $j=1$ the bracket vanishes and the degree-one count is $2\sum_a(b_1(a)-b_2(a))$. Summing over $j$, the two sums of shifted $A_i$ terms cancel since $A_0=0$, leaving $2(2^{|W|}-1)$ functions in this family. In \eqref{rtwo:balancedproduct}, choosing $t$ adds $v_p(n-t)-v_p(t)$ to the exponent of $X_p$ for each prime $p$. These sums all vanish precisely when $D_S=1$, by unique factorization. Negative exponents are allowed, and $A_0=0$ excludes the empty set.

The exceptional functions in Theorem~\ref{rtwo:turns} occur only from dimension six onwards. By symmetry their zero sets reduce to $Z=\{1,n\}$ or $\{0,1,n\}$. For a singleton accepting weight, the least degree is $|Z|$ by the root bound and a normalized product. In even dimension $n=2m$, the only possible pair of accepting weights in this orientation is $m,m+1$. The factor $(t-1)(t-n)$ takes the same nonzero value at both, giving degree two for the two-weight set $Z$. Multiplication by $t$ makes the values unequal, so for the three-weight set $Z$ the interpolant needs a linear factor and the function has degree four. In even dimension, the singleton choices contribute eight functions of degree two and eight of degree three after reflection and output complementation. The pair contributes four more of degree two and four of degree four. In odd dimension only a singleton is possible, giving four each of degrees two and three. These are the terms $e_{n,j}$. Equation~\eqref{ra:sampling} rules out three changes for a two-query test, and a constant costs zero queries.
\end{proof}

\section{The second counting term at degree three}\label{cub:sec}
At degree three, four specified weights account for the leading count in Theorem~\ref{dg:all}. The first correction comes from supports of sizes five and six. Four-weight labellings whose degree drops contribute only to the $O(N^3)$ remainder. Write $I_N=\{0,\ldots,N-1\}$ with $N=n+1$, and let $C_j$ count degree-three functions with exactly $j$ specified weights. The six-weight contribution comes from the classical count on Vinogradov's quadric, where two triples in $I_N$ have the same sum and the same sum of squares. In the five-weight count, interpolation determines a sixth integer coordinate, which may fall outside $I_N$. For example, $t(t-4)(t-5)/12$ vanishes at $0,4,5$ and equals one at $1,2,6$. The labels at $0,1,2,4,5$ therefore admit a sixth weight for $n=6$, but already exhaust their cubic's Boolean values on the grid for $n=5$.

\begin{proposition}[Cubic count and extension from five to six weights]\leavevmode\label{cub:second}%
Let $N=n+1$ and $K_5=2+2\pi/(3\sqrt3)$. As $N\to\infty$, the number of symmetric partial Boolean functions of free real degree three is
\begin{equation}\label{cub:expansion}
 14\binom N4+
 \left(\frac{5}{2\pi^2}+\frac{2}{3\sqrt3\,\pi}\right)N^3\log N+O(N^3).
\end{equation}
The $N^3\log N$ term comes from the counts on five and six specified weights, respectively,
\begin{equation}\label{cub:supportcounts}
 C_5=\frac{K_5}{\pi^2}N^3\log N+O(N^3), C_6=\frac{1}{2\pi^2}N^3\log N+O(N^3).
\end{equation}
Exactly $6C_6$ of the $C_5$ functions admit a sixth weight while retaining degree three. Thus the extendible fraction, whenever $C_5>0$, is $6C_6/C_5$, which equals $3/K_5+O(1/\log N)$ as $N\to\infty$.
\end{proposition}

\begin{proof}\phantomsection\label{pr:cub:second}
For a cubic $p$, each of $p$ and $p-1$ has at most three roots, while three or fewer specified weights admit a quadratic interpolant. Thus only support sizes four, five and six occur. On four weights all $14$ nonconstant labellings have degree at most three. Subtracting the quadratic exceptions in Proposition~\ref{dg:finite} gives $C_4=14\binom N4-2\sum_{a\ge1}\binom{N-2a}{2}=14\binom N4+O(N^3)$.

On five weights the label multiplicities must be three and two. By exchanging labels we may take the three zeros to be $x_1,x_2,x_3$ and the two ones to be $y_1,y_2$. Every representing cubic is proportional to $P(t)=\prod_i(t-x_i)$, so a cubic exists exactly when $P(y_1)=P(y_2)$. The common value is nonzero because the five specified weights are distinct. The monic cubic $P(t)-P(y_1)$ already has roots $y_1,y_2$. Its quadratic coefficient determines the third root $y_3=\sum_i x_i-y_1-y_2$, an integer which may lie outside $I_N$. The resulting two triples satisfy
\begin{equation}\label{cub:moments}
 \sum_{i=1}^3x_i=\sum_{i=1}^3y_i,\qquad
 \sum_{i=1}^3x_i^2=\sum_{i=1}^3y_i^2.
\end{equation}
Conversely, the first identity gives equal quadratic coefficients for the two monic cubics. Together the identities give equal linear coefficients, since $2\sum_{i<j}x_ix_j=(\sum_i x_i)^2-\sum_i x_i^2$, and likewise for $y$. The cubics therefore differ by a constant. In particular $P(y_1)=P(y_2)$ and $P(t)/P(y_1)$ realizes the five labels.

Let $J_5$ count ordered solutions of \eqref{cub:moments} with $x_1,x_2,x_3,y_1,y_2\in I_N$ and $y_3\in\mathbb Z$, and let $J_6$ impose the grid constraint also on $y_3$. Even the unrestricted coordinate is $O(N)$ by the first identity. Solutions with a repeated entry contribute $O(N^3)$. Indeed, removing a shared entry leaves two pairs with the same sum and product, hence the same entries up to permutation. If a triple has a repeated value, that value, its remaining entry and one entry of the other triple determine the remaining pair up to permutation.

For five specified weights, discard solutions with a repetition among the five specified coordinates. The computed $y_3$ may still equal $y_1$ or $y_2$. For example, with zeros $0,5,8$ and ones $2,9$, we have $P(t)-36=(t-2)^2(t-9)$, so the only candidate is the already specified weight two. Let $J_5^*$ count ordered solutions with five distinct specified coordinates. For six specified weights, let $J_6^*$ count solutions with all six coordinates distinct. The exact conversions are
\begin{equation}\label{cub:orderedcounts}
 C_5=\frac{2J_5^*}{3!\,2!}=\frac{J_5^*}{6},C_6=\frac{J_6^*}{3!\,3!}=\frac{J_6^*}{36}.
\end{equation}
In the first count either label may be the majority, and only two minority weights are specified. In the second count we assign the first triple to the zeros. The repetition bound gives $J_i^*=J_i+O(N^3)$ for $i=5,6$. Rogovskaya obtained $J_6=(18/\pi^2)N^3\log N+O(N^3)$~\cite{cubrogo,cubkeil}. Blomer and Br\"udern give a finer expansion~\cite{cubblomer}.

It remains to compute $J_5$ with one coordinate unrestricted. Average over the six choices of this coordinate. Each choice has the same count by permutations within triples and interchange of the triples. Outside $O(N^3)$ solutions the three differences $x_i-y_i$ are nonzero. Their sum is zero, so there are six sign patterns. Once the unrestricted coordinate is summed over, simultaneous permutations and interchange of the triples give the same total for each pattern. We may thus keep the pattern $(+,+,-)$ and sum over the six choices of unrestricted coordinate without the averaging factor $1/6$.

Write the differences in this pattern as $(gu,gv,-gh)$, where $g,u,v>0$, $\gcd(u,v)=1$ and $h=u+v$. Here $(u,v)$ specifies the direction and $g$ its scale.

Fix a direction $(u,v)$. Translating all coordinates by $-y_3$ preserves both moment equalities. Set $a=u/h$, $b=v/h$. The sum-of-squares equation is now
$u(y_1-y_3)+v(y_2-y_3)=-g(u^2+uv+v^2)$.
A particular solution for the differences is $(-hg,-vg)$. Coprimality gives all solutions by adding $m(v,-u)$, $m\in\mathbb Z$. With $t=m/g$, the translated coordinates in the order $(x_1,x_2,x_3,y_1,y_2,y_3)$, divided by $hg$, are
\begin{equation}\label{cub:coordinates}
 b(t-1),\quad -at,\quad -1,\quad bt-1,\quad -at-b,\quad 0.
\end{equation}
For unrestricted coordinate $i$, let $W_i(a,t)$ be the largest minus the smallest of the other five entries. Their unscaled span is $hgW_i(a,t)$. A set of integer span $D$ has exactly $\max\{0,N-D\}$ translates inside $I_N$, so for fixed $u,v,i$ the contribution is
\begin{equation}\label{cub:lattice}
 \sum_{g\ge1,\,m\in\mathbb Z}\max\{0,N-hgW_i(a,m/g)\}
  =\frac{N^3}{6h^2}\int_{\mathbb R}\frac{dt}{W_i(a,t)^2}
    +O(N^2/h+N).
\end{equation}
To see that the error is uniform in the direction, write $gW_i(a,m/g)$ as the range of five linear forms in $(g,m)$. This extends the range to $g=0$ and gives a uniform Lipschitz bound, since the coefficients are bounded. If the range vanishes, five forms agree. Equality of the triple sums then forces all six to agree. In particular $0,-g,b(m-g),-am$ agree, giving $g=m=0$ since $a+b=1$. By compactness the range has a positive lower bound on the unit semicircle, uniformly in $a\in[0,1]$ and $i$.

After scaling $(g,m)$ by $h/N$ and dividing the summand by $N$, we therefore have a uniformly Lipschitz function of bounded support. Applying the Lipschitz bound on interior mesh cells and including the boundary cells at $g=0$ gives the error in \eqref{cub:lattice}. To evaluate the integral, put $m=gt$. The Jacobian is $g$, and the inner integral is $\int_0^{1/W}g(1-gW)\,dg=1/(6W^2)$ with $W=W_i(a,t)$. The support bound restricts the contributing directions to $h\le cN$ for an absolute $c$. There are at most $h$ directions at each $h$, so their errors total $O(N^3)$.

We integrate the inverse squares of the spans and then sum over directions. For the six entries in~\eqref{cub:coordinates}, the range on $t\le0$, $0\le t\le1$ and $t\ge1$ is $1-t,1,t$, respectively. Its inverse square integrates to three. At almost every $t$, deleting any of the four nonextreme entries leaves that span unchanged, contributing $4\cdot3=12$ to the sum of the six integrals.

After deleting the maximum, split each interval where the two candidates for the new span $W$ agree. Table~\ref{tab:cubspans} gives the resulting spans and split points.
\begin{table}[H]
\caption{Spans and integrals after deletion of the maximum.}\label{tab:cubspans}
\centering\small
\begin{tabular}{@{}llll@{}}\toprule
Interval & Span $W$ & Split point & $(1-ab)(L,R)$\\ \midrule
$t\le0$ & $\max(a-t,1-bt)$ & $-b/a$ & $(a,b)$\\
$0\le t\le1$ & $\max(1-at,a+bt)$ & $b$ & $(b,a)$\\
$t\ge1$ & $\max(b+at,t-a)$ & $1/b$ & $(a,b)$\\ \bottomrule
\end{tabular}
\par\smallskip
\begin{minipage}{.90\linewidth}\footnotesize
$L$ and $R$ are the integrals of $W^{-2}$ to the left and right of the split point within the indicated interval. The last column gives both integrals multiplied by $1-ab$.
\end{minipage}
\end{table}
For instance, on $t\le-b/a$ the span is $a-t$, giving $L=1/(a+b/a)=a/(1-ab)$. Since $a+b=1$, each row contributes $1/(1-ab)$. Replacing $t$ by $1-t$ negates the list and translates it by $-1$, up to permutation. Deleting the minimum therefore gives the same integral as deleting the maximum. Since $1-ab=1-a+a^2$, the sum over all six deletions is
\begin{equation}\label{cub:spankernel}
 K(a):=\sum_{i=1}^6\int_{\mathbb R}W_i(a,t)^{-2}\,dt
       =12+\frac{6}{1-a+a^2}.
\end{equation}
To sum over primitive directions, use $\gcd(u,v)=\gcd(u,h)=1$. The explicit formula for $K$ in~\eqref{cub:spankernel} gives a Lipschitz bound on $[0,1]$, so $\sum_{1\le v<j}K(v/j)=j\int_0^1K(a)\,da+O(1)$.
Writing $\mathbf1_{\{\gcd(u,h)=1\}}=\sum_{q\mid\gcd(u,h)}\mu(q)$, where $\mu$ is the M\"obius function, and applying this Riemann sum gives
\begin{equation}\label{cub:directions}
 \sum_{h\le X}\frac1{h^2}\sum_{\substack{1\le u<h\\\gcd(u,h)=1}} K(u/h)
   =\frac{\log X}{\zeta(2)}\int_0^1K(a)\,da+O(1).
\end{equation}
Indeed, writing $h=qj$ after inversion gives the outer weights $\mu(q)/q^2$. The inner harmonic sum is $\log(X/q)+O(1)$, and the total error is bounded since $\sum_{q\ge1}(1+\log q)/q^2$ converges. Now $\zeta(2)=\pi^2/6$ and $\frac16\int_0^1 K(a)\,da=2+\int_0^1(1-a+a^2)^{-1}\,da=2+2\pi/(3\sqrt3)=K_5$. Applying \eqref{cub:directions} with $X=cN$ to \eqref{cub:lattice} gives $J_5=(K_5/\zeta(2))N^3\log N+O(N^3)$. By~\eqref{cub:orderedcounts} and the repetition bound, dividing $J_5$ by six and $J_6$ by 36 gives the stated coefficients of $C_5$ and $C_6$. Adding $C_4$ yields \eqref{cub:expansion}.

Finally, count extendible five-weight labellings by deleting one weight from a six-weight labelling. Each of the six deletions retains degree three, since a majority triple together with an opposite label excludes degree two. Conversely, a five-weight labelling determines the cubic. The cubic already attains the majority value at three distinct weights, so a sixth weight would have to carry the minority label at the unique candidate $y_3$. An extension exists exactly when this candidate is new and belongs to $I_N$. Thus deletion is a bijection between a six-weight labelling with one marked weight and an extendible five-weight labelling. Substituting the counts in~\eqref{cub:supportcounts} gives
$6C_6/C_5=6[1/(2\pi^2)+O(1/\log N)]/[K_5/\pi^2+O(1/\log N)]$, as required.
\end{proof}

\end{appendices}

\end{document}